\documentclass[english,15pt]{article}
\usepackage[T1]{fontenc}
\usepackage[latin1]{inputenc}
\usepackage{geometry}
\usepackage{float}
\usepackage{mathrsfs}
\usepackage{amsmath}
\usepackage{amssymb}
\usepackage{graphicx}
\usepackage{hyperref}
\usepackage[color=yellow]{todonotes}
\hypersetup{
    colorlinks=true,
    linkcolor=blue,
    filecolor=magenta,      
    urlcolor=cyan,
    citecolor = blue,
}
\usepackage{yhmath}
\makeatletter
\usepackage{bm}
\numberwithin{equation}{section}
\usepackage{array}

\usepackage[english]{babel}
\floatstyle{ruled}
\usepackage{algorithm}
 \usepackage{algorithmicx}

\usepackage{amsthm}

\usepackage{mathrsfs}

\usepackage{amsfonts}

\usepackage{epsfig}

\usepackage{bm}

\usepackage{mathrsfs}

\usepackage{enumerate}

\@ifundefined{definecolor}{\@ifundefined{definecolor}
 {\@ifundefined{definecolor}
 {\usepackage{color}}{}
}{}
}{}
\newtheorem{cor}{Corollary}[section]

\usepackage{subcaption}

\newtheorem{theorem}{Theorem}[section]
\newtheorem{lem}{Lemma}[section]
\newtheorem{rem}{Remark}[section]
\newtheorem{proposition}{Proposition}[section]
\newcounter{hypA}

\newcounter{hypB}

\newcounter{hypD}

\usepackage{babel}\date{}

\newcommand{\EE}{\mathbb{E}}

\newcommand{\Pa}{ {\cal P }}

\def \EE{\mathbb{E}}

\usepackage{mathtools}

\numberwithin{equation}{section}

\makeatother

\begin{document}

\begin{center}

{\Large \textbf{Multilevel Localized Ensemble Kalman--Bucy Filters}}

\vspace{0.5cm}

 NEIL K. CHADA$^{\ddagger}$

{\footnotesize $^{\dagger}$Department of Mathematics
, City University of Hong Kong, \\ 83 Tat Chee Ave, Kowloon Tong, Hong Kong SAR.}  \\
{\footnotesize E-Mail:\,}
\texttt{\emph{\footnotesize neilchada123@gmail.com}} \\
\begin{abstract}
\textcolor{black}{In this article we propose and develop a new methodology which is inspired from Kalman filtering and multilevel Monte Carlo (MLMC), entitle the multilevel localized ensemble Kalman--Bucy Filter (MLLEnKBF). Based on the work of Chada et al. \cite{CJY20}, we provide an important extension on this which is to include the technique of covariance localization. Localization is important as it can induce stability and remove long spurious correlations, particularly with a small ensemble size. Our resulting algorithm is used for both state and parameter estimation, for the later we exploit our method for normalizing constant estimation. As of yet, MLMC has only been applied to localized data assimilation methods in a discrete-time setting, therefore this work acts as a first in the continuous-time setting.  Numerical results indicate its performance, and benefit through a range of model problems, which include a linear Ornstein--Uhlenbeck process, of moderately high dimension, and the Lorenz 96 model, for parameter estimation. Our results demonstrate improved stability, and that with MLMC, one can reduce the computational complexity to attain an order is MSE $\mathcal{O}(\epsilon^2)$, for $\epsilon>0$.} 
\end{abstract}
\end{center}

\section{Introduction}

Data assimilation (DA) \cite{BC09,CR11,AJ70} is an important and relevant topic of interest within applied mathematics and statistics. It is concerned with the culmination of data and model dynamics, to improve on model prediction and inference. Traditionally DA arose in control theory which was concerned with applications in signal processing, and also dynamical systems. For signal processing, and other aspects of control theory, it was where Rudolph Kalman developed an approximation scheme for data assimilation, which is known as the Kalman filter (KF). The KF, as stated in its original paper, is concerned with the updating of a state, through data, with is done through the mean and covariance. This is possible when one assumes a linear and Gaussian system. This is of relevance as in that particular setup the optimal filter, which minimizes the difference between the truth and approximate filter is indeed the Kalman filter. Our focus in this paper is on the continuous-time setting where our two governing equations take the form
\begin{align}
\label{eq:data}
dY_t & =  CX_t dt + R_2^{1/2} dV_t, \\
\label{eq:signal}
dX_t & =  A X_t dt + R_1^{1/2} dW_t,
\end{align}
where \eqref{eq:data} $\{Y_t\}_{t \geq 0}$ denotes the observational process, and \eqref{eq:signal} $\{X_t\}_{t \geq 0}$ denotes the signal process. The exact interpretations of the constants appearing in both equations will be explained later, but we will state now that $\{V_t\}_{t \geq 0}$ and $\{W_t\}_{t \geq 0}$ represent $d$-dimensional Brownian motion, where noise is added to both equations. 
In particular we are interested in the filtering problem, which is to compute  conditional expectation $\mathbb{E}[\varphi(X_t)|\mathscr{F}_t]$, 
 where $\varphi:\mathbb{R}^{d_x} \rightarrow \mathbb{R}$  is an appropriately integrable function and $\{\mathscr{F}_t\}_{t \geq 0}$ is the filtration generated by the observed process \eqref{eq:data}.
 
 Since the formulation of the KF, many extensions have been considered, with arguably the most successful of these being the ensemble Kalman filter (EnKF) \cite{GE09,GE94}, in discrete-time. The EnKF differs to the KF in that it updates an ensemble of particles using sample means and covariances. Thus resulting in a more computationally cheaper method. It can be viewed as a Monte-Carlo version of the KF. Since its formulation, it has been applied in a number of different applications such as numerical weather prediction, geophysical science and medical imaging \cite{MW06,ORL08}. However to solve the filtering problem related to \eqref{eq:data}-\eqref{eq:signal}, we require the use of continuous-time Kalman methods such as the Kalman Bucy filter (KBF) and the ensemble Kalman Bucy filter (EnKBF) \cite{BD20} (which is the continuous-time version of the EnKF).

\begin{figure}[h!]
\centering
\includegraphics[width=0.55\textwidth]{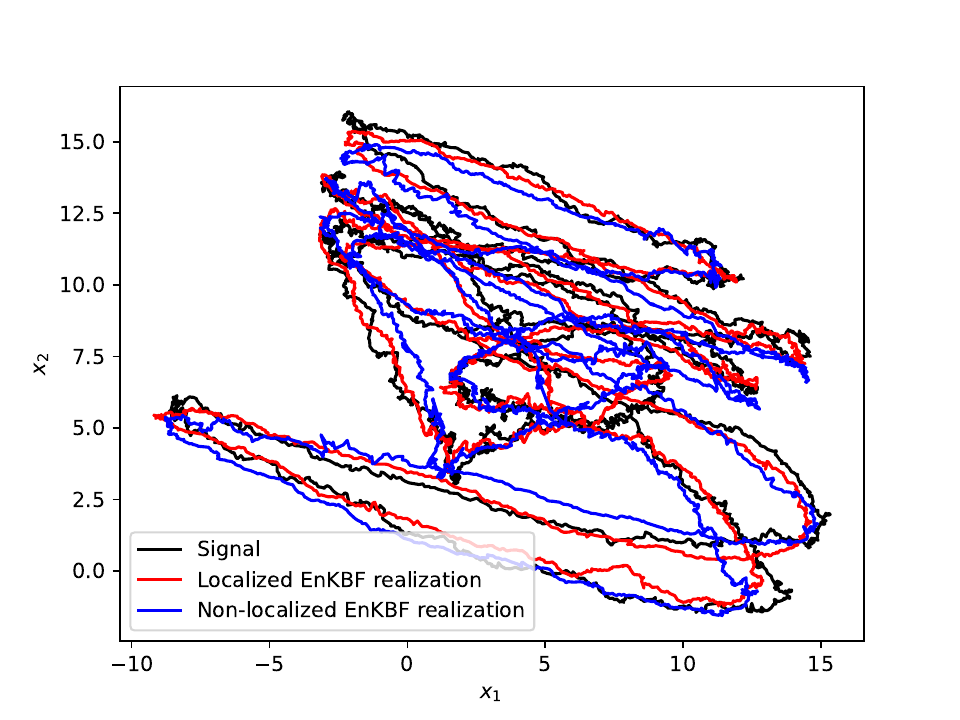}
 \caption{Numerical simulation of a localized and non-localized EnKBF aiming to recover the signal, described through Equation \eqref{eq:signal}.}
 \label{fig:local_benefits}
\end{figure}

Despite this, there are certain limitations that can occur within the EnKF, where numerous remedies have been developed. Such issues include long spurious correlation, and the performance of the EnKF with a small ensemble size, which can tend to under-represent the state or parameter. To overcome such problems there are commonly two remedies: the first is variance inflation, which inflates the covariance by a noise term. The second, which is important and relevant to this work, is localization \cite{JLA07,BR10,HWS01,OHS04}. Localization is a technique designed to remove these issues, and operates by considering a ``localized" covariance function, which is a Schur product of the covariance and a localization function.
The benefits of localization are that in induces stability for problems when the EnKF is applied to the dynamical systems. This is evident from Figure \ref{fig:local_benefits} where we provide a comparison of the EnKF and EnKF with localization on the continuous-dynamics of \eqref{eq:data}-\eqref{eq:signal}.
The benefits of such a technique have been documented well, and is almost seen as a ``\textit{gold-standard}" when implementing the EnKF. As a result there has been considerable attraction to understanding its theoretical aspects \cite{XTT18,TM23}. This includes work from Reich, Wiljes et al. \cite{WRS18} and also the extension to the continuous-time setting, resulting in a localized EnKBF. However aside from this, there has been limited work done in the continuous setting \cite{WT20}. Therefore the focus, and aim of this paper are to continue in this direction when we aim to develop a multilevel Monte Carlo (MLMC) estimator related to the LEnKBF. MLMC \cite{MBG08,MBG15} is a popular variance reduction technique, used to reduce the cost associated to attaining a particular order of MSE $\mathcal{O}(\epsilon^2)$, for $\epsilon>0$. Recently this methodology has been applied to a range of data assimilation methods such as the particle filter, sequential Monte Carlo sampler, and both the EnKF and EnKBF. However as of yet there has not been a synergy in MLMC and localization techniques for continuous-time methods. The only existing work, to the best of our knowledge, is the work of Fossum et al. \cite{FT17,NBF22} which consider this in the discrete-time setting for specific problems arising in geophysical sciences. As a result we propose a new algorithm which is a multilevel localized ensemble Kalman--Bucy filter (MLLEnKBF). By considering a localized version of the MLEnKBF, we are able to show improvements related to stability, which can become apparent in the non-localized cases. Therefore our motivation and contribution is primarily from a computational standpoint. As the original MLEnKBF was developed for a linear and Gaussian problem, we continue this here where we test this on an Ornstein-Uhlenbeck process. As developing any theory is beyond the scope of this work, we leave that for future endeavors. To reinstate we are focused on the computational aspect. To complement our work we also consider the application of the MLLEnKBF for normalizing constant estimation (NCE). This is based on the work of Ruzaquat et al. \cite{RCJ21} which proposed how one can use the MLEnKBF for NCE. Estimating the NC is of use as it has benefits in traditional statistics such as Bayes Factor for hypothesis testing, and for estimating full filtering distributions. We extend this slightly for both linear and nonlinear models. For the linear we consider the ML complexity rates, while for the nonlinear examples we will exploit our algorithm for parameter estimation.

\subsection{Outline}
The outline of this paper is as follows: In Section \ref{sec:model} we provide the necessary background material related to the Kalman-Bucy filter, and the Ensemble Kalman-Bucy filter and its connections with MLMC. We then discuss localization, specifically covariance localization in Section \ref{sec:local} and present our main algorithm of interest for the paper, the MLEnKBF. Numerical simulations are provided in Section \ref{sec:num}, on a number of model problems. We conclude our findings in Section \ref{sec:conc} where we also present some future directions of research.

\section{Model problem}
\label{sec:model}

In this section we provide an overview of the necessary background material required for the rest of the article. We begin by providing the continuous-time filtering problem through the Kalman--Bucy filters (KBF). This will lead onto a discussion of the ensemble Kalman--Bucy filter (EnKBF), which is an $N-$particle representation of the KBF. Finally we will discuss the concept of multilevel Monte Carlo (MLMC), and review its extension to the EnKBF, referred to as the MLEnKBF. 
\subsection{Kalman--Bucy Filters}

Consider a linear-Gaussian filtering model of the following form
\begin{align}
\label{eq:data}
dY_t & =  CX_t dt + R_2^{1/2} dV_t, \\
\label{eq:signal}
dX_t & =  A X_t dt + R_1^{1/2} dW_t,
\end{align}
where $(Y_t,X_t)\in\mathbb{R}^{d_y}\times\mathbb{R}^{d_x}$, $(V_t,W_t)$ is a $(d_y+d_x)-$dimensional standard Brownian motion, 
$A$ is a square $d_x\times d_x$ matrix, $C$ is a $d_y\times d_x$ matrix, $Y_0=0$, $X_0\sim\mathcal{N}_{d_x}(\mathcal{M}_0,\mathcal{P}_0)$
($d_x-$dimensional Gaussian distribution, mean $\mathcal{M}_0$, covariance matrix $\mathcal{P}_0$)
and $R_1^{1/2},R_2^{1/2}$ are square (of the appropriate dimension) and symmetric and invertible matrices. It is well-known that, letting
$\{\mathscr{F}_t\}_{t\geq 0}$ be the filtration generated by the observations, the conditional probability of $X_t$ given $\mathscr{F}_t$ is a
Gaussian distribution with mean and covariance matrix 
$$
\mathcal{M}_t := \mathbb{E}[X_t|\mathscr{F}_t], \quad \mathcal{P}_t := \mathbb{E}\Big([X_t - \mathbb{E}(X_t|\mathscr{F}_t)][X_t - \mathbb{E}(X_t|\mathscr{F}_t)]^{\top}\Big),
$$
given by the Kalman--Bucy and Ricatti equations \cite{DT18}
\begin{align}
\label{eq:kbf}
d\mathcal{M}_t &= A \mathcal{M}_t dt + \mathcal{P}_tC^{\top}R^{-1}_2\Big(dY_t - C\mathcal{M}_t dt\Big), \\
\label{eq:ricc}
\partial_t\mathcal{P}_t &= \textrm{Ricc}(\mathcal{P}_t),
\end{align}
where the Riccati drift term is defined as
$$
\textrm{Ricc}(Q) = AQ + QA^{\top}-QSQ + R, \quad \textrm{with} \ R =R_1 \quad \textrm{and} \ S:=C^{\top}R^{-1}_2C.
$$
 A derivation of \eqref{eq:kbf} - \eqref{eq:ricc} can be found in \cite{AJ70}. The KBF is viewed as the $\mathbb{L}_2$-optimal state estimator for an Ornstein--Uhlenbeck process, given the state is partially observed with linear and Gaussian assumptions. The Kalman--Bucy diffusion is a conditional Mckean-Vlasov diffusion process.
An alternative approach is to consider a conditional nonlinear McKean-Vlasov type diffusion process. For this article we work with three different processes of the form
\begin{eqnarray}
d\overline{X}_t & = &A~\overline{X}_t~dt~+~R^{1/2}_{1}~d\overline{W}_t+\Pa_{\eta_t}C^{\top}R^{-1}_{2}~\left[dY_t-\left(C\overline{X}_tdt+R^{1/2}_{2}~d\overline{V}_{t}\right)\right],\label{eq:vanilla_kbf}\\
d\overline{X}_t & = &A~\overline{X}_t~dt~+~R^{1/2}_{1}~d\overline{W}_t+\Pa_{\eta_t}C^{\top}R^{-1}_{2}~\left[dY_t-\left(\frac{1}{2}C\left[\overline{X}_t+\eta_t(e)\right]dt\right)\right],
\label{eq:determ_kbf}\\
d\overline{X}_t & = &A~\overline{X}_t~dt~+~R_1\Pa_{\eta_t}^{-1}\left(\overline{X}_t-\eta_t(e)\right)~dt+\Pa_{\eta_t}C^{\top}R^{-1}_{2}~\left[dY_t-\left(\frac{1}{2}C\left[\overline{X}_t+\eta_t(e)\right]dt\right)\right],
\label{eq:determ_transp_kbf}
\end{eqnarray}

where $(\overline{V}_t,\overline{W}_t,\overline{X}_0)$ are independent copies of $(V_t,W_t,X_0)$ and covariance
$$
\mathcal{P}_{\eta_t} = \eta_t\Big([e-\eta_t(e)][e-\eta_t(e)]^{\top}\Big), \quad \eta_t:= \mathrm{Law}(\overline{X}_t|\mathscr{F}_t),
$$
such that $\eta_t$ is the conditional law of $\overline{X}_t$ given $\mathscr{F}_t$ and $e(x)=x$. We will explain the difference of each diffusion process, in succeeding subsections. It is important to note that the nonlinearity in \eqref{eq:vanilla_kbf}-\eqref{eq:determ_transp_kbf} does not depend on the distribution of the state $\mathrm{Law}(\overline{X}_t)$ but on the conditional distribution $\eta_t$, and $\mathcal{P}_{\eta_t}$ alone does not depend on $\mathscr{F}_t$. These processes are
commonly referred to as Kalman-Bucy (nonlinear) diffusion processes.
It is known that the conditional  expectations of the random states $\overline{X}_t$ and their conditional covariance matrices $\mathcal{P}_{\eta_t}$, w.r.t. $\mathscr{F}_t$, satisfy the Kalman--Bucy and the Riccati equations. In addition, for any $t\in\mathbb{R}^+$
$$
\eta_t:= \mathrm{Law}(\overline{X}_t|\mathscr{F}_t) = \mathrm{Law}({X}_t|\mathscr{F}_t).
$$
As a result, an alternative to recursively computing \eqref{eq:kbf} - \eqref{eq:ricc}, is to generate $N$ i.i.d.~samples from any of
\eqref{eq:vanilla_kbf}-\eqref{eq:determ_transp_kbf} processes and apply a Monte Carlo approximation, as mentioned which we now discuss.

\subsection{Ensemble Kalman--Bucy Filter}

Exact simulation from \eqref{eq:vanilla_kbf}-\eqref{eq:determ_transp_kbf} is typically not possible, or feasible, as one cannot compute $\mathcal{P}_{\eta_t}$ exactly. 
The ensemble Kalman--Bucy filter (EnKBF) can be used to deal with this issue. The EnKBF coincides with the mean-field particle interpretation of the Kalman-Bucy diffusion processes. The EnKBF is an $N-$particle system that is simulated as follows, for the $i^{th}-$particle, $i\in\{1,\dots,N\}$:
\begin{eqnarray}
d\xi_t^i & = &A~\xi_t^i~dt~+~R^{1/2}_{1}~d\overline{W}^i_t+P_t^NC^{\top}R^{-1}_{2}~\left[dY_t-\left(C\xi_t^idt+R^{1/2}_{2}~d\overline{V}^i_{t}\right)\right],
\label{eq:enkbf}\\
d\xi_t^i & = &A~\xi_t^i~dt~+~R^{1/2}_{1}~d\overline{W}^i_t+P_t^NC^{\top}R^{-1}_{2}~\left[dY_t-\left(\frac{1}{2}C\left[\xi_t^i+ \mathcal{M}_t \right]dt\right)\right],
\label{eq:denkbf}\\
d\xi_t^i & = &A~\xi_t^i~dt~+~R_1 (P_t^N)^{-1}\left(\xi_t^i-\mathcal{M}_t\right)~dt+P_t^N C^{\top}R^{-1}_{2}~\left[dY_t-\left(\frac{1}{2}C\left[\xi_t^i+\mathcal{M}_t\right]dt\right)\right],
\label{eq:dtenkbf}
\end{eqnarray}
such that the sample mean and covariances are defined as
\begin{align*}
{P}_t^N & =  \frac{1}{N -1}\sum_{i=1}^N (\xi_t^i-m^N_t)(\xi_t^i-m^N_t)^{\top},\\
m^N_t & =  \frac{1}{N }\sum_{i=1}^N \xi_t^i,
\end{align*}
where $\xi_0^i\stackrel{\textrm{i.i.d.}}{\sim}\mathcal{N}_{d_x}(\mathcal{M}_0,\mathcal{P}_0)$. It is remarked that when $C=0$, \eqref{eq:enkbf} and \eqref{eq:denkbf} reduce to $N$--independent copies of an Ornstein--Uhlenbeck process. 

The first EnKBF, described through \eqref{eq:enkbf}, is known as the vanilla EnKBF (VEnKBF), which is the standard EnKBF used in theory and practice which contains perturbed observations, through the brownian motion $\overline{V}_t$. Equation \eqref{eq:denkbf}, is referred to as the deterministic EnKBF (DEnKBF), which, unlike \eqref{eq:enkbf}, contains no perturbed observations. Finally  \eqref{eq:dtenkbf} is known as the deterministic transport EnKBF (DTEnKBF), where the modification is that it does not contain not containing $\overline{W}_t$, implying it is completely deterministic. 

In practice, one will not have access to an entire trajectory of observations. Thus numerically, one often works with a time discretization, such as an Euler-type discretization.
Let $\Delta_l=2^{-l}$ denote our level of discretization, then we will generate the system for $(i,k)\in\{1,\dots,N\}\times\mathbb{N}_0=\mathbb{N}\cup\{0\}$ as


\begin{eqnarray}
   \xi_{(k+1)\Delta_l}^i & = & \xi_{k\Delta_l}^i + A  \xi_{k\Delta_l}^i  \Delta_l + R_1^{1/2} \big\{\overline{W}_{(k+1)\Delta_l}^i-\overline{W}_{k\Delta_l}^i \big\} + \nonumber \\
   \label{eq:enkbf1}
 & & P^N_{k\Delta_l} C^{\top} R_2^{-1} \Big( \big\{Y_{(k+1)\Delta_l} - Y_{k\Delta_l} \big\} - \Big[ C \xi_{k\Delta_l}^i  \Delta_l + R_2^{1/2} \big \{\overline{V}_{(k+1)\Delta_l}^i - \overline{V}_{k\Delta_l}^i \big\} \Big] \Big),\\
 \xi_{(k+1)\Delta_l}^i & =&  \xi_{k\Delta_l}^i + A  \xi_{k\Delta_l}^i  \Delta_l + R_1^{1/2} \big\{\overline{W}_{(k+1)\Delta_l}^i-\overline{W}_{k\Delta_l}^i \big\} + \nonumber \\
    \label{eq:denkbf1}
 & & P^N_{k\Delta_l} C^{\top} R_2^{-1} \left( \big\{Y_{(k+1)\Delta_l} - Y_{k\Delta_l} \big\} - C \left(\dfrac{ \xi_{k\Delta_l}^i + m^N_{k\Delta_l}}{2} \right)\Delta_l \right),\\
 \xi_{(k+1)\Delta_l}^i & = & \xi_{k\Delta_l}^i + A  \xi_{k\Delta_l}^i  \Delta_l + R_1 \left(P_{k\Delta_l}\right)^{-1} \left(\xi_{k\Delta_l}^i - m^N_{k\Delta_l}  \right) \Delta_l + \nonumber\\
     \label{eq:dtenkbf1}
 & & P^N_{k\Delta_l} C^{\top} R_2^{-1} \left( \big\{Y_{(k+1)\Delta_l} - Y_{k\Delta_l} \big\} - C \left(\dfrac{ \xi_{k\Delta_l}^i + m^N_{k\Delta_l}}{2} \right)\Delta_l \right),
\end{eqnarray}


such that
\begin{align*}
P_{k\Delta_l}^N & =  \frac{1}{N -1}\sum_{i=1}^N (\xi_{k\Delta_l}^i-m_{k\Delta_l^N})(\xi_{k\Delta_l}^i-m_{k\Delta_l}^N)^{\top}, \\
m_{k\Delta_l}^N & =  \frac{1}{N }\sum_{i=1}^N \xi_{k\Delta_l}^i,
\end{align*}
and $\xi_0^i\stackrel{\textrm{i.i.d.}}{\sim}\mathcal{N}_{d_x}(\mathcal{M}_0,\mathcal{P}_0)$. For $l\in\mathbb{N}_0$ given, denote by $\eta_t^{N,l}$ as the $N-$empirical
probability measure of the particles $(\xi_t^1,\dots,\xi_t^N)$, where $t\in\{0,\Delta_l,2\Delta_l,\dots\}$. For $\varphi:\mathbb{R}^{d_x}\rightarrow\mathbb{R}^{d_x}$
we will use the notation $\eta_t^{N,l}(\varphi):=\tfrac{1}{N}\sum_{i=1}^N\varphi(\xi_t^{i})$.

\subsection{Multilevel EnKBF}

Let us define $\pi$ to be be a probability on a measurable space $(\mathsf{X},\mathscr{X})$ and for $\pi-$integrable $\varphi:\mathsf{X}\rightarrow\mathbb{R}$
consider the problem of estimating $\pi(\varphi)=\mathbb{E}_{\pi}[\varphi(X)]$. Now let us assume that we only have access to a sequence of approximations of $\pi$,  $\{\pi_l\}_{l\in\mathbb{N}_0}$, also each defined on $(\mathsf{X},\mathscr{X})$ and we are now interested in estimating $\pi_l(\varphi)$, such that  $\lim_{l\rightarrow\infty}|[\pi_l-\pi](\varphi)|=0$. Therefore one can use the telescoping sum
$$
\pi_L(\varphi) = \pi_0(\varphi) + \sum^L_{l=1}[\pi_l-\pi_{l-1}](\varphi),
$$
as we know that the approximation error between $\pi$ and $\pi_l$ gets smaller as $l \rightarrow \infty$. The idea of multilevel Monte Carlo (MLMC) is to construct a coupled system, related to the telscoping sum, such that the mean sqaured error can be reduced,  relative to i.i.d.~sampling from $\pi_L$. Therefore our MLMC approximation of $\mathbb{E}_{\pi_L}[\varphi(X)]$ is 
$$
\pi_L^{ML}(\varphi) := \frac{1}{N_0}\sum_{i=1}^{N_0}\varphi(X^{i,0}) + \sum_{l=1}^L \frac{1}{N_l}\sum_{i=1}^{N_l}\{\varphi(X^{i,l})-\varphi(\tilde{X}^{i,l-1})\},
$$
where $N_0\in\mathbb{N}$ i.i.d.~samples from $\pi_0$ as $(X^{1,0},\dots,X^{N_0,0})$ and for $l\in\{1,\dots,L\}$, $N_l\in\mathbb{N}$ samples
from a coupling of $(\pi_l,\pi_{l-1})$ as $((X^{1,l},\tilde{X}^{1,l-1}),\dots,(X^{N_l,l},\tilde{X}^{N_l,l-1}))$.
Then the MSE is then
$$
\mathbb{E}[(\pi_L^{ML}(\varphi)-\pi(\varphi))^2] = \underbrace{\mathbb{V}\textrm{ar}[\pi_L^{ML}(\varphi)]}_{\textrm{variance}} + 
[\underbrace{\pi_L(\varphi)-\pi(\varphi)}_{\textrm{bias}}]^2.
$$
The work of Chada et al. \cite{CJY20} proposed the application of MLMC for the EnKBF, where we will now briefly review this. If we consider the discretized vanilla EnKBF \eqref{eq:enkbf1}, then, for  $l\in\mathbb{N}$ and $(i,k)\in\{1,\dots,N\}\times\mathbb{N}_0$ the ML adaption is given as

\[
\textbf{(F1)}
\begin{cases}
\begin{aligned}
\xi_{(k+1)\Delta_l}^{i,l} & = \xi_{k\Delta_l}^{i,l} + A\xi_{k\Delta_l}^{i,l}\Delta_l + R_1^{1/2} [\overline{W}_{(k+1)\Delta_l}^i-\overline{W}_{k\Delta_l}^i]  \\
&+ P_{k\Delta_l}^{N,l}C^{\top}R_2^{-1}\Big([Y^{{i}}_{(k+1)\Delta_l}-Y^{{i}}_{k\Delta_l}]-\Big[C\xi_{k\Delta_l}^{i,l}\Delta_l + R_2^{1/2}[\overline{V}_{(k+1)\Delta_l}^i-\overline{V}_{k\Delta_l}^i]\Big]\Big),  \\
\xi_{(k+1)\Delta_{l-1}}^{i,l-1} & =  \xi_{k\Delta_{l-1}}^{i,l-1} + A\xi_{k\Delta_{l-1}}^{i,l-1}\Delta_{l-1} + R_1^{1/2} [\overline{W}_{(k+1)\Delta_{l-1}}^i-\overline{W}_{k\Delta_{l-1}}^i] \\
 &+ P_{k\Delta_{l-1}}^{N,l-1}C^{\top}R_2^{-1}\Big([Y^{{i}}_{(k+1)\Delta_{l-1}}-Y^{{i}}_{k\Delta_{l-1}}] -\Big[C\xi_{k\Delta_{l-1}}^{i,l-1}\Delta_{l-1}+ R_2^{1/2}[\overline{V}_{(k+1)\Delta_{l-1}}^i-\overline{V}_{k\Delta_{l-1}}^i]\Big]\Big).
\end{aligned}
\end{cases}
\] \\
 Similarly for the deterministic variant \eqref{eq:denkbf1} we have
\[
\textbf{(F2)}
\begin{cases}
\begin{aligned} 
\xi_{(k+1)\Delta_l}^{i,l} & = \xi_{k\Delta_l}^{i,l} + A\xi_{k\Delta_l}^{i,l}\Delta_l + R_1^{1/2} [\overline{W}_{(k+1)\Delta_l}^i-\overline{W}_{k\Delta_l}^i] \\
&+ P_{k\Delta_l}^{N,l}C^{\top}R_2^{-1}\left([Y^{{i}}_{(k+1)\Delta_l}-Y^{{i}}_{k\Delta_l}]-C \left(\dfrac{ \xi_{k\Delta_l}^{i,1} + m^{N,l}_{k\Delta_l}}{2} \right)\Delta_l \right),  \\
\xi_{(k+1)\Delta_{l-1}}^{i,l-1} & =  \xi_{k\Delta_{l-1}}^{i,l-1} + A\xi_{k\Delta_{l-1}}^{i,l-1}\Delta_{l-1} + R_1^{1/2} [\overline{W}_{(k+1)\Delta_{l-1}}^i-\overline{W}_{k\Delta_{l-1}}^i] \\
 &+ P_{k\Delta_{l-1}}^{N,l-1}C^{\top}R_2^{-1}\left([Y^{{i}}_{(k+1)\Delta_{l-1}}-Y^{{i}}_{k\Delta_{l-1}}] -C \left(\dfrac{ \xi_{k\Delta_{l-1}}^{i,l-1} + m^{N,l-1}_{k\Delta_{l-1}}}{2} \right)\Delta_{l-1} \right),
\end{aligned}
\end{cases}
\] \\
where our sample covariances and means are defined accordingly as
\begin{align*}
P_{k\Delta_l}^{N,l} & =  \frac{1}{N -1}\sum_{i=1}^N (\xi_{k\Delta_l}^{i,l}-m_{k\Delta_l}^{N,l})(\xi_{k\Delta_l}^{i,l}-m_{k\Delta_l}^{N,l})^{\top} ,\\
m_{k\Delta_l}^{N,l} & =  \frac{1}{N }\sum_{i=1}^N \xi_{k\Delta_l}^{i,l},\\
P_{k\Delta_{l-1}}^{N,l-1} & =  \frac{1}{N -1}\sum_{i=1}^N (\xi_{k\Delta_{l-1}}^{i,l-1}-m_{k\Delta_{l-1}}^{N,l-1})(\xi_{k\Delta_{l-1}}^{i,l-1}-m_{k\Delta_{l-1}}^{N,l-1})^{\top}, \\
m_{k\Delta_{l-1}}^{N,l-1} & =  \frac{1}{N }\sum_{i=1}^N \xi_{k\Delta_{l-1}}^{i,l-1},
\end{align*}
and $\xi_0^{i,l}\stackrel{\textrm{i.i.d.}}{\sim}\mathcal{N}_{d_x}(\mathcal{M}_0,\mathcal{P}_0)$, $\xi_{0}^{i,l-1}=\xi_0^{i,l}$.
Then, one has the approximation of $[\eta_t^l-\eta_t^{l-1}](\varphi)$, $t\in\mathbb{N}_0$, $\varphi:\mathbb{R}^{d_x}\rightarrow\mathbb{R}$, given as
$$
[\eta_t^{N,l} -\eta_t^{N,l-1}](\varphi) = \frac{1}{N}\sum_{i=1}^N[\varphi(\xi_t^{i,l})-\varphi(\xi_t^{i,l-1})].
$$
Therefore for the multilevel estimation, one has an approximation for $t\in\mathbb{N}_0$
\begin{equation}\label{eq:main_est}
\eta_t^{ML}(\varphi):=\eta_t^{N_0,0}(\varphi) + \sum_{l-1}^L [\eta_t^{N_l,l} -\eta_t^{N_l,l-1}](\varphi).
\end{equation}
Given the multilevel estimator, the following Theorem below characterizes the MSE-to-cost rate, related
to the MLEnKBF. We highlight for convenience that cost is related to the functional evaluations, and not
CPU/clock-time. The theorem translates to that in order to achieve an MSE of order $\mathcal{O}(\epsilon^2)$,
then the cost associated with it, is of order $\mathcal{O}(\epsilon^{-2}\log(\epsilon)^2)$. This is an improvement
over the single-level EnKBF where the cost to achieve the same MSE is of order $\mathcal{O}(\epsilon^3)$. This was proven
for the case of $\textbf{(F1)}$ in \cite{ACJ22} and $\textbf{(F2)}$ in \cite{ACJ22}. We highlight it is still an open problem to show this for $\textbf{(F3)}$, thus why we do not consider this variant within the rest of the manuscript.

\begin{theorem}
\label{theo:main_theo}
For any $T\in\mathbb{N}$ fixed and $t\in[0,T-1]$ there exists a $\mathsf{C}<+\infty$ such that for any $(L,N_{0:L})\in\mathbb{N}\times\{2,3,\dots\}^{L+1}$,
$$
\mathbb{E}\left[\left\|[\hat{\eta}_t^{ML}-\eta_t](\varphi)\right\|_2^2\right] \leq \mathsf{C}\left(
\sum_{l=0}^L \frac{\Delta_l}{N_l} + \sum_{l=1}^L\sum_{q=1, q\neq l}^L\frac{\Delta_l\Delta_q}{N_lN_q} + \Delta_L^2
\right).
$$
\end{theorem}

\subsection{Normalizing constant estimation}
We now turn our attention to the computation of the normalization constant, associated with the filtering distribution.
This will prove useful later on, we adapt our methodologies to parameter estimation. The normalizing constant is an important
quantity in statistics, also referred to as  the marginal likelihood. It is particularly useful
as it can be used for model comparison, which can include mixture models such as time-series state space models, 
and hierarchical models.
\\
We begin by defining the normalizing constant, of the filtering distribution,  as 
$$
Z_t:=\frac{{\mathcal L}_{X_{0:t},Y_{0:t}}}{{\mathcal L}_{X_{0:t},W_{0:t}}},
$$ to be the density of ${\mathcal L}_{X_{0:t},Y_{0:t}}$, the law of the process $(X,Y)$ and that of ${\mathcal L}_{X_{0:t},W_{0:t}}$, the law of the process $(X,W)$. That is, 
\[
{\mathbb E}[f(X_{0:t})g(Y_{0:t})]= {\mathbb E}[f(X_{0:t})g(W_{0:t})Z_{t}(X,Y)].
\]
One can show that 
\begin{equation}
\label{eq:initt}
Z_t(X,Y)=\exp{\left[\int_0^t \left[\langle CX_s, R_2^{-1} dY_s\rangle -\frac{1}{2}\langle X_s,SX_s\rangle ~ds\right]\right]},
\end{equation}
where  $S:=C^{\top}R_2^{-1}C$. 
Now we let $\overline{Z}_t(Y)$ denote the likelihood function defined by
$$
\overline{Z}_t(Y):=\EE_Y\left(Z_t(X,Y)\right),
$$
where $\EE_Y\left(\cdot\right)$ stands for the expectation w.r.t. the signal process when the observation is fixed and independent of the signal.
From the work of Crisan et al. \cite{CDJ21}, the authors show that the normalizing constant is given by
\begin{align}
\label{eq:nc}
\overline{Z}_t(Y)=\exp{\left[\int_0^t 
\left[~\langle C\mathcal{M}_s, R_2^{-1} dY_s\rangle -\frac{1}{2}\langle \mathcal{M}_s,S\mathcal{M}_s\rangle~ds\right]
\right]},
\end{align}
and a suitable estimator of it is defined as
\begin{align}
\label{eq:nc_estimate}
\overline{Z}_t^N(Y)=\exp{\left[\int_0^t 
\left[~\langle Cm_s^N, R_2^{-1} dY_s\rangle -\frac{1}{2}\langle m_s^N,Sm_s^N\rangle~ds\right]
\right]},
\end{align}
which follows from replacing the conditional mean of the signal process  with the sample mean associated with the EnKBF. 
\begin{rem}
    To briefly describe how \eqref{eq:nc} was attained in \cite{CDJ21}, the authors used a change of measure rule, through Girsanov's Theorem, then an application of Bayes' Theorem which can be expressed through as the Kallianpur-Striebel formula defined on path space. The derivation is a standard approach, which is discussed in  Chapter 3 in \cite{BC09}.
\end{rem}
In practice, one must time-discretize the EnKBF, and the normalizing constant estimator \eqref{eq:nc_estimate} to yield for $t\in\mathbb{N}$ 
\begin{align}
\label{eq:disc_nc_estimate}
\overline{Z}_t^{N,l}(Y)=\exp\Bigg\{\sum_{k=0}^{t\Delta_l^{-1}-1} 
\left[~\langle Cm_{k\Delta_l}^N, R_2^{-1} [Y_{(k+1)\Delta_l}-Y_{k\Delta_l}]\rangle -\frac{\Delta_l}{2}\langle m_{k\Delta_l}^N,Sm_{k\Delta_l}^N\rangle\right]
\Bigg\}. 
\end{align}
Let $\overline{U}_t^{N,l}(Y)=\log(\overline{Z}_t^{N,l}(Y))$, where we now consider the estimation of log-normalization constants. To enhance the efficiency we consider a coupled ensemble Kalman--Bucy filter, as described in Section \ref{sec:model}. Let $l\in\mathbb{N}$ then we run the coupled system of the different MLEnKBFs. 
As before with Theorem \ref{theo:main} we have an analgous theorem in the context of MLEnKBFs for NC estimation,
which is proven and taken from \cite{RCJ21}

\begin{theorem}
\label{theo:main}
For any $T\in\mathbb{N}$ fixed and $t\in[0,T-1]$ there exists a $\mathsf{C}<+\infty$ such that for any $(L,N_{0:L})\in\mathbb{N}\times\{2,3,\dots\}^{L+1}$,
$$
\mathbb{E}\left[\left\|[\widehat{\overline{U}}_t^{ML}-{\overline{U}}_t](Y)\right\|_2^2\right] \leq \mathsf{C}\left(
\sum_{l=0}^L \frac{\Delta_l}{N_l} + \sum_{l=1}^L\sum_{q=1, q\neq l}^L\frac{\Delta_l\Delta_q}{N_lN_q} + \Delta_L^2
\right).
$$
The above theorem translates as, in order to achieve an MSE of order $\mathcal{O}(\epsilon^2)$, for $\epsilon>0$, we have a cost of $\mathcal{O}(\epsilon^{-2} ~\log(\epsilon)^2)$. This implies a reduction in cost compared to the single-level NC estimator.
\end{theorem}

\section{Localization \& Multilevel Localized EnKBF}
\label{sec:local}

In this section we introduce and discuss localization and its relevance in the context of
ensemble Kalman filtering. We will then introduce our numerical scheme entitle the localized multilevel ensemble Kalman filter, and describe some its properties before explaining it in
algorithmic form. This will also lead onto the normalizing constant estimator that we will use
later in the numerical experiments. We emphasize again our form of localization we consider is
covariance localization. Other forms exists such as domain localization, which is is not the focus in this work.
\\
Let us assume we have a fixed ensemble size $N$ within our ensemble Kalman method. These methods
are known to encapsulate much oft he informative dynamics when the size of $N$ is low. However this can cause
problems related to undersampling. When the ensemble size $N$ is small,
it is not able to provide an accurate statistical representative of the state of a system. 
As a result undersampling can lead to inbreeding, as well as filter divergence, and eventually spurious
correlations which are long range. It is possible to implement counter measures, where one these ways
is through covariance localization. it is a process of cutting off longer range correlations in the error covariances at a specified distance. While doing so it improves the estimate of the forecast
error covariance. It operates by applying a Schur product. 

\subsection{Localization functions}
The simple modification of the Localized EnKBF (e.g.~\cite{WT20}) is simply to replace $P_t^N$ in \eqref{eq:enkbf}-\eqref{eq:dtenkbf} (or $P_{k\Delta_l}^N$ in \eqref{eq:enkbf1}-\eqref{eq:dtenkbf1}) by a localized version:
$P_t^{N,\textrm{loc}}$ with $(P_t^{N,\textrm{loc}})_{ij}=(P_t^{N})_{ij}\phi_{ij}$, where $\phi$ is a localization matrix which essentially removes dependence between co-ordinates that are far apart; see \cite{WT20} for details. This localization can, under assumptions, provide an $\mathcal{O}(1)$ cost for a fixed ensemble size as the dimension increases. Moreover this localized EnKBF can be trivially expanded to the MLEnKBF (at least for the stochastic ones) and to estimate normalizing constants.

To achieve this one has to apply a  Schur product with a function $\rho$ which we define as our localization function.
There are multiple choices of localization functions where that one can commonly choose from. Below we provide some of these. Before discussing this we introduce some notation.
Our localization function is denoted as $\phi_r(d):\mathbb{R}^+_{\geq 0} \rightarrow [0,1]$, where $r>0$ denotes the localization radius, and $d \in \mathbb{R}^+$ is the dimension, 
which must satisfy the following
$$
\phi_r(0) = 1, \quad \phi_r(d)=0, \quad \forall d, \quad r>0.
$$
Three commonly chosen choices of localization functions are presented below:
\begin{itemize}
\item \textit{Uniform localization function}: $\phi_r(d) = 1_{[0,1]}(d)$.
\item \textit{Triangular localization function}: $\phi_r(d) = (1-\frac{d}{r}) 1_{[0,1]}(d)$.
\item \textit{Gaspari-Cohn localization function}: 
\[
\phi_r(d) = 
\begin{cases}
\begin{aligned}
-8\frac{d^5}{r^5} + 8\frac{d^4}{r^4} + 5\frac{d^3}{r^3} - \frac{20}{3} \frac{d^2}{r^2} + 1, \qquad 0 \leq d < \frac{r}{2}, \\
\frac{8}{3}\frac{d^5}{r^5} - 8\frac{d^4}{r^4} + 5\frac{d^3}{r^3} + \frac{20}{3} \frac{d^2}{r^2} - 10\frac{r}{d} + 4, \qquad \frac{r}{2} \leq d < 1.
\end{aligned}
\end{cases}
\] \\
\end{itemize}
In the above functions, $d$  is the Euclidean distance between either the grid points in physical space or the
distance between a grid point and the observation location. Furthermore $r \in \mathbb{R}^+$ is an important parameter, which
represents s defined such that beyond this the
correlation reduces from 1.0 and at a distance of more than twice the correlation length
scale the correlation reduces to zero. Commonly the correlation is chosen as $r=10/3l$, where $l$ is a cut-off length.
\\
To help visualize the different localization functions, we plot them in Figure \ref{fig:local_funct}.
An important question related to the functions stated above is, what is a suitable choice. Based on this
question we will use the Gaspari-Cohn (GC) localization function. The reasons for this is that it has
been demonstrated the perform well when used as the localization function, and the
localization is a function of only the horizontal distance between an observation and a state
variable. Due to this, Tuning the localization only requires finding an appropriate value for the parameter
that defines the half-width. Given the computational gains with localization it has been successful in 
applications within the use of Kalman methods \cite{RJS15,XTT18,TM23}. However most of this has been on the discrete-time
EnkF, where for us we are concerned with the continuous-time setting, i.e. the EnKBF. 

Thus far there has been limited work in this direction with the most notable work by De Wiljes et al. \cite{WT20}, which considers the analysis for a localized EnKBF with complete observations. We now introduce our main algorithm of interest which is the MLEnKBF, which we present for the cases of both $\textbf{F(1)}$ and $\textbf{F(2)}$, i.e.

\[
\textbf{(L-F1)}
\begin{cases}
\begin{aligned}
\xi_{(k+1)\Delta_l}^{i,l} & = \xi_{k\Delta_l}^{i,l} + A\xi_{k\Delta_l}^{i,l}\Delta_l + R_1^{1/2} [\overline{W}_{(k+1)\Delta_l}^i-\overline{W}_{k\Delta_l}^i]  \\
&+ P_{k\Delta_l}^{\textrm{loc},N,l}C^{\top}R_2^{-1}\Big([Y^{{i}}_{(k+1)\Delta_l}-Y^{{i}}_{k\Delta_l}]-\Big[C\xi_{k\Delta_l}^{i,l}\Delta_l + R_2^{1/2}[\overline{V}_{(k+1)\Delta_l}^i-\overline{V}_{k\Delta_l}^i]\Big]\Big),  \\
\xi_{(k+1)\Delta_{l-1}}^{i,l-1} & =  \xi_{k\Delta_{l-1}}^{i,l-1} + A\xi_{k\Delta_{l-1}}^{i,l-1}\Delta_{l-1} + R_1^{1/2} [\overline{W}_{(k+1)\Delta_{l-1}}^i-\overline{W}_{k\Delta_{l-1}}^i] \\
 &+ P_{k\Delta_{l-1}}^{\textrm{loc},N,l-1}C^{\top}R_2^{-1}\Big([Y^{{i}}_{(k+1)\Delta_{l-1}}-Y^{{i}}_{k\Delta_{l-1}}] -\Big[C\xi_{k\Delta_{l-1}}^{i,l-1}\Delta_{l-1}+ R_2^{1/2}[\overline{V}_{(k+1)\Delta_{l-1}}^i-\overline{V}_{k\Delta_{l-1}}^i]\Big]\Big).
\end{aligned}
\end{cases}
\] \\
 Similarly for the deterministic variant \eqref{eq:denkbf1} we have
\[
\textbf{(L-F2)}
\begin{cases}
\begin{aligned} 
\xi_{(k+1)\Delta_l}^{i,l} & = \xi_{k\Delta_l}^{i,l} + A\xi_{k\Delta_l}^{i,l}\Delta_l + R_1^{1/2} [\overline{W}_{(k+1)\Delta_l}^i-\overline{W}_{k\Delta_l}^i] \\
&+ P_{k\Delta_l}^{\textrm{loc},N,l}C^{\top}R_2^{-1}\left([Y^{{i}}_{(k+1)\Delta_l}-Y^{{i}}_{k\Delta_l}]-C \left(\dfrac{ \xi_{k\Delta_l}^{i,1} + m^{N,l}_{k\Delta_l}}{2} \right)\Delta_l \right),  \\
\xi_{(k+1)\Delta_{l-1}}^{i,l-1} & =  \xi_{k\Delta_{l-1}}^{i,l-1} + A\xi_{k\Delta_{l-1}}^{i,l-1}\Delta_{l-1} + R_1^{1/2} [\overline{W}_{(k+1)\Delta_{l-1}}^i-\overline{W}_{k\Delta_{l-1}}^i] \\
 &+ P_{k\Delta_{l-1}}^{\textrm{loc},N,l-1}C^{\top}R_2^{-1}\left([Y^{{i}}_{(k+1)\Delta_{l-1}}-Y^{{i}}_{k\Delta_{l-1}}] -C \left(\dfrac{ \xi_{k\Delta_{l-1}}^{i,l-1} + m^{N,l-1}_{k\Delta_{l-1}}}{2} \right)\Delta{l-1} \right),
\end{aligned}
\end{cases}
\] 
where now the key difference is with our covariance function $P^{\textrm{loc}}$ in discretized form.
Our contributed algorithms are now provided in Algorithm \ref{alg:MLEnKBF}, and Algorithm \ref{alg:MLEnKBF_NC} for the normalizing constant estimation. 
\begin{rem}
    We remark that in order for us to derive MSE-to-cost rates, one requires a ``true" solution. In the linear-case the EnKBF converges to the KBF, however this is not the case for the localized version hence we cannot proceed with this in the analysis, however for the numerical simulation's careful consideration is required to handle this. We will defer this discussion until the next section.
\end{rem}
\begin{figure}[h!]
\centering
\hspace*{-2cm}
\includegraphics[width=0.5\textwidth]{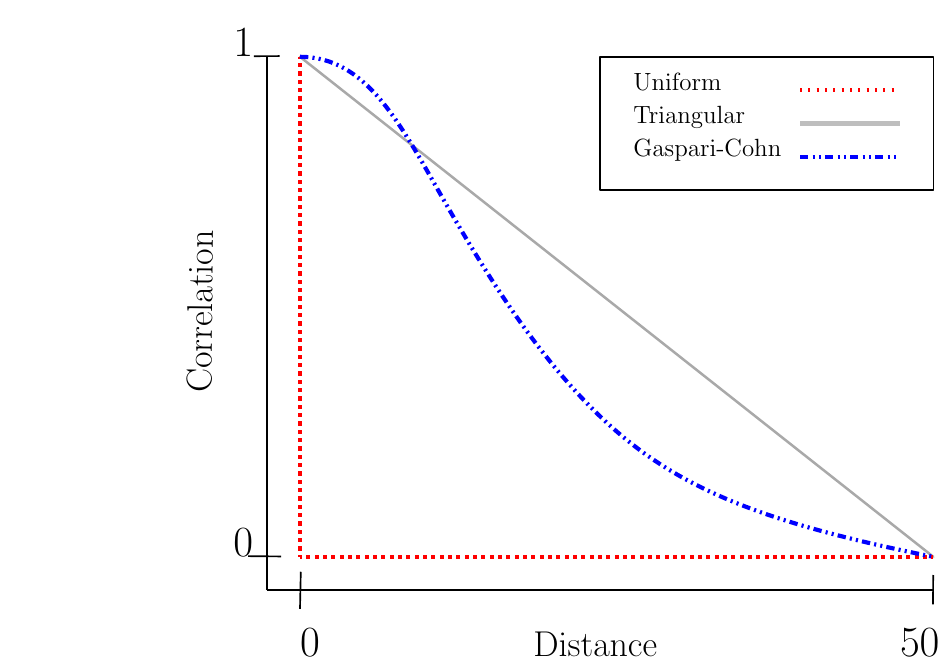}
 \caption{Graphical representation of the different localization functions discussed. They include the uniform (red), the triangular (grey) and the Gaspari-Cohn (blue)
 localization functions.}
 \label{fig:local_funct}
\end{figure}

\begin{algorithm}[h!]
\caption{(\textbf{LMLEnKBF}) Localized Multilevel Ensemble Kalman--Bucy Filter}
\label{alg:MLEnKBF}
\begin{enumerate}
\item \textbf{Input:} Target level $L\in\mathbb{N}$, start level $l_*\in \mathbb{N}$ such that $l_*<L$, the number of particles on each level $\{N_l\}_{l=l_*}^L$, the time parameter $T\in\mathbb{N}$ and initial independent ensembles $\Big\{\{\tilde{\xi}_0^{i,l_*}\}_{i=1}^{N_{l_*}}, \cdots, \{\tilde{\xi}_0^{i,L}\}_{i=1}^{N_{L}}\Big\}$.
\item \textbf{Initialize:} Set $l=l_*$. For $(i,k)\in \{1,\cdots,N_l\}\times \{0,\cdots,T\Delta_l^{-1}-1\}$, set $\{\xi_0^{i,l}\}_{i=1}^{N_l} = \{\tilde{\xi}_0^{i,l}\}_{i=1}^{N_l}$ . Then return $\eta_T^{N_l,l}(\varphi)$.
\item \textbf{Iterate:} For $l \in \{l_*+1,\cdots,L\}$ and $(i,k)\in \{1,\cdots,N_l\} \times \{0,\cdots,T\Delta_l^{-1}-1\}$, set $\{\xi_0^{i,l-1}\}_{i=1}^{N_l}=\{\xi_0^{i,l}\}_{i=1}^{N_l}= \{\tilde{\xi}_0^{i,l}\}_{i=1}^{N_l}$ (the one which corresponds to the case used in Step 2.). Then using \textbf{(L-F1)}-\textbf{(L-F2)}, return $\eta_T^{N_l,l-1}(\varphi)$ \& $\eta_T^{N_l,l}(\varphi)$.
\item \textbf{Output:} Return the multilevel estimation of the L-EnKBF:
\begin{align}
\label{eq:MLEnKBF_NC}
\eta_{T,\textrm{loc}}^{ML}(\varphi) = \eta_{T,\textrm{loc}}^{N_{l_*},l_*}(\varphi) + \sum_{l=l_*+1}^L \{\eta_{T,\textrm{loc}}^{N_l,l}(\varphi)-\eta_{T,\textrm{loc}}^{N_l,l-1}(\varphi)\}.
\end{align}
\end{enumerate}
\end{algorithm}

\begin{algorithm}[h!]
\caption{\textcolor{black}{(\textbf{LMLEnKBF-NC})} Localized Multilevel Estimation of Normalizing Constants}
\label{alg:MLEnKBF_NC}
\begin{enumerate}
\item \textbf{Input:} Target level $L\in\mathbb{N}$, start level $l_*\in \mathbb{N}$ such that $l_*<L$, the number of particles on each level $\{N_l\}_{l=l_*}^L$, the time parameter $T\in\mathbb{N}$ and initial independent ensembles $\Big\{\{\tilde{\xi}_0^{i,l_*}\}_{i=1}^{N_{l_*}}, \cdots, \{\tilde{\xi}_0^{i,L}\}_{i=1}^{N_{L}}\Big\}$.
\item \textbf{Initialize:} Set $l=l_*$. For $(i,k)\in \{1,\cdots,N_l\}\times \{0,\cdots,T\Delta_l^{-1}-1\}$, set $\{\xi_0^{i,l}\}_{i=1}^{N_l} = \{\tilde{\xi}_0^{i,l}\}_{i=1}^{N_l}$ and simulate any of the cases \eqref{eq:enkbf1}-\eqref{eq:dtenkbf1} to return $\{m^{N_l,l}_{k\Delta_l}\}_{k=0}^{T\Delta_l^{-1}-1}$. Then using \eqref{eq:disc_nc_estimate}, return $\overline{Z}_T^{N_l,l}(Y)$ or $\overline{U}_T^{N_l,l}(Y)$.
\item \textbf{Iterate:} For $l \in \{l_*+1,\cdots,L\}$ and $(i,k)\in \{1,\cdots,N_l\} \times \{0,\cdots,T\Delta_l^{-1}-1\}$, set $\{\xi_0^{i,l-1}\}_{i=1}^{N_l}=\{\xi_0^{i,l}\}_{i=1}^{N_l}= \{\tilde{\xi}_0^{i,l}\}_{i=1}^{N_l}$, and simulate any of the coupled ensembles \textbf{(L-F1)}-\textbf{(L-F2)} (the one which corresponds to the case used in Step 2.) to return $\{m^{N_l,l-1}_{k\Delta_{l-1}}\}_{k=0}^{T\Delta_{l-1}^{-1}-1}$ and $\{m^{N_l,l}_{k\Delta_l}\}_{k=0}^{T\Delta_l^{-1}-1}$. Then using \eqref{eq:disc_nc_estimate}, return $\overline{Z}_T^{N_l,l-1}(Y)$ \& $\overline{Z}_T^{N_l,l}(Y)$ or $\overline{U}_T^{N_l,l-1}(Y)$ \& $\overline{U}_T^{N_l,l}(Y)$.
\item \textbf{Output:} Return the multilevel estimation of the normalizing constant:
\begin{align}
\label{eq:MLEnKBF_NC}
\overline{Z}_{T,\text{loc}}^{ML}(Y) = \overline{Z}_T^{N_{l_*},l_*}(Y) + \sum_{l=l_*+1}^L \{\overline{Z}_T^{N_l,l}(Y)-\overline{Z}_T^{N_l,l-1}(Y)\}.
\end{align}
or its logarithm:
\begin{align}
\label{eq:MLEnKBF_Log_NC}
\overline{U}_{T,\text{loc}}^{ML}(Y) = \overline{U}_T^{N_{l_*},l_*}(Y) + \sum_{l=l_*+1}^L \{\overline{U}_T^{N_l,l}(Y)-\overline{U}_T^{N_l,l-1}(Y)\}.
\end{align}
\end{enumerate}
\end{algorithm}

\newpage

In order for us to proceed with the numerical experiments we require a particular bound on the variance of the difference
of our estimators, at different discretization levels. This is to provide a sanity check for us in the numerical experiments. We will verify this result both analytically and computationally. This is analogous to the assumption we make in \cite{ACJ22}. The result is presented as follows.

\begin{lem}[Strong error]\label{lem:strong_error}
For any $T\in\mathbb{N}$ fixed and $t\in[0,T-1]$ there exists a $\mathsf{C}<+\infty$ such that for any $(l,j,k_1)\in\mathbb{N}_0\times\{1,\dots,d_x\}\times\{0,1,\dots,\Delta_{l}^{-1}\}$:
$$
\mathbb{E}\Big[\Big(\overline{X}_{t+k_1\Delta_{l}}(j)-\overline{X}_{t+k_1\Delta_{l}}^l(j)\Big)^2\Big] \leq \mathsf{C}\Delta_l^2.
$$
\end{lem}

\begin{proposition}[Variance bound]
\label{prop:main}
Given a set of discretization levels $l \in \{1,\ldots,L\}$ and corresponding amount of particles $\{N_l\}_{l=1}^L$,
the following bound on the incremental difference between level $l$ and $l-1$ holds
$$
\mathbb{E} \big[ \| \Xi \|^2_2\big] \leq \mathsf{C} \Delta_l,
$$
for $C>0$ and $\Delta_l = 2^{-l}>0$. 
\end{proposition}

\begin{proof}
We begin our proof by stating the $C_p$ inequality which is given as
$$
\mathbb{E}[|X+Y|]^p \leq C_p [\mathbb{E}[|X|]^p + \mathbb{E}[|Y|]^p], \quad q \geq 2,
$$
which we require. 
Now we proceed
\begin{align*}
\mathbb{E} \big[ \| \Xi \|^2_2\big] &= \mathbb{E}\Big[ \|\eta^{N_l,l}_t - \eta^{N_{l-1},l-1}_t \|^2_2 \Big] \\
&=  \mathbb{E}\Big[ \| \varphi(\xi^{i,l}_t) -  \varphi(\xi^{i,l}_t)\|^2_2 \Big] \\
&=  \mathbb{E}\Big[ \| \varphi(\xi^{i,l}_t) + \varphi(\xi_t) - \varphi(\xi_t) -  \varphi(\xi^{i,l-1}_t)\|^2_2 \Big] \\
&\leq \frac{1}{N_l}\sum^L_{l-1}\sum^{N}_{i=1} \mathbb{E}\Big[ \|\varphi(\xi^{i,l}_t) + \varphi(\xi_t) - \varphi(\xi_t) -  \varphi(\xi^{i,l-1}_t)\|^2_2 \Big] \\
& \leq \frac{1}{N_l}\sum^L_{l-1}\sum^{N}_{i=1} {C}_2 \Bigg(\mathbb{E}\Big[\|\varphi(\xi^{i,l}_t) - \varphi(\xi_t)\|^2_2\Big] + \mathbb{E}\Big[\| \varphi(\xi_t)- \varphi(\xi^{i,l-1}_t)\|^2_2\Big]\Bigg).
\end{align*}
Then using the weak, or strong, rate of convergence results from Lemma \ref{lem:strong_error},
we have that
\begin{align*}
\mathbb{E}\Big[ \|\eta^{N_l,l}_t - \eta^{N_{l-1},l-1}_t \|^2_2 \Big] &\leq \frac{1}{N_l}\sum^L_{l-1}\sum^{N}_{i=1} {C}_2\Big(C_3\Delta_l^2)^{1/2} + C_4(\Delta_l^2)^{1/2}\Big) \\
& \leq \mathsf{C}\Delta_l,
\end{align*}
which completes the proof.
\end{proof}
The above result holds in the case of the non-localized, we now provide a direct corollary for the localized version.
\begin{cor}
Given a set of discretization levels $l \in \{1,\ldots,L\}$ and corresponding amount of particles w $\{N_l\}_{l=1}^L$,
the following bound on the incremental difference between level $l$ and $l-1$ holds
$$
\mathbb{E} \big[ \| \Xi_{\mathrm{loc}} \|^2_2\big] \leq \mathsf{C} \Delta_l,
$$
for $C>0$ and $\Delta_l = 2^{-l}>0$. 
\end{cor}
\begin{proof}
The proof follows directly from Theorem \ref{prop:main}, with the strong error remaining the same with the inclusion of a localized covariance matrix.
\end{proof}

The above results state the variance should decay similarly at the same rate, which we can observe in Figure \ref{fig:local_funct}. The key difference that we observe, is that the variance is considerably lower for localization, which is what we believed before. The difference in variance is approximately of an order $\mathcal{O}(10^1).$

\begin{figure}[h!]
\centering
\includegraphics[width=0.55\textwidth]{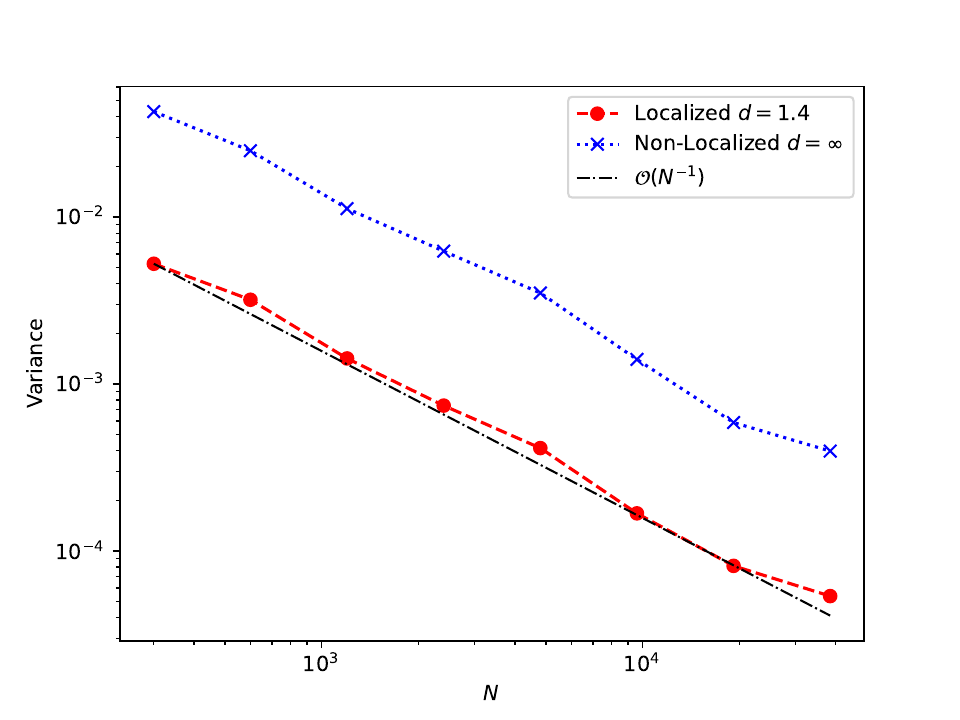}
 \caption{variance plot of the respective MSE's related to the problem that we know.}
 \label{fig:local_funct}
\end{figure}

Finally, we are now in a position to conjecture a result, which is analogous to Theorem \ref{theo:main_theo}, where we use a change in notation to denote our localized estimator.

\begin{theorem}{(Conjectured)}
\label{theo:local_main}
For any $T\in\mathbb{N}$ fixed and $t\in[0,T-1]$ there exists a $\mathsf{C}<+\infty$ such that for any $(L,N_{0:L})\in\mathbb{N}\times\{2,3,\dots\}^{L+1}$,
$$
\mathbb{E}\left[\left\|[\hat{\eta}_{t,\text{loc}}^{ML}-\eta_t](\varphi)\right\|_2^2\right] \leq \mathsf{C}\left(
\sum_{l=0}^L \frac{\Delta_l}{N_l} + \sum_{l=1}^L\sum_{q=1, q\neq l}^L\frac{\Delta_l\Delta_q}{N_lN_q} + \Delta_L^2
\right).
$$
The above theorem translates as, in order to achieve an MSE of order $\mathcal{O}(\epsilon^2)$, for $\epsilon>0$, we have a cost of $\mathcal{O}(\epsilon^{-2} ~\log(\epsilon)^2)$. This implies a reduction in cost compared to the single-level NC estimator.
\end{theorem}

\begin{rem}
We highlight a number of points in this remark. Firstly that the above theorem is related to both the vanilla and deterministic counterparts. Secondly, (ii) it does not only hold for localized MLEnKBF estimators, but also the normalizing constant estimation. Finally (iii), as stated we conjecture this results holds true. In other words we expect to see a rate of $\mathcal{O}(\epsilon^{-2} ~\log(\epsilon)^2)$. We hypothesis this, as the additional analysis goes beyond the work of this paper. Our aim for future work is to fully quantify this. We again note our focus here is on the computational developments.
\end{rem}
\section{Numerical Experiments}
\label{sec:num}

In this section we provide our highlighting contributions, which are the numerical simulations of our newly proposed localized multilevel ensemble Kalman--Bucy filter. We will consider a number of numerical experiments, where firstly we provide a verification of the hypothesized theorem, in the section above, related to the rate of complexity. This will be compared to the non-localized multilevel schemes, where we aim to demonstrate the with the reduction in variance, we can aim to achieve a lower MSE for the same computational burden. This will be verified also for the normalizing constant estimator. Finally we present a prameter estimation example where we use the Lorenz-96 Model to learn the parameter of interest,




\subsection{State estimation}

We begin this discussion with verifying our hypothesis of the MSE-to-cost multilevel rates, which we believe to be also the same for the normalizing constants. As the analysis, and computation in \cite{CJY20,RCJ21} holds only in the linear case, we consider only this model. Throughout our experiments, we will assume that the state space dimnesion is the same as the observational space dimension, i.e. $d_x=d_y$.

\subsubsection{Linear Gaussian model}

\begin{figure}[h!]
\centering
\includegraphics[width=0.49\textwidth]{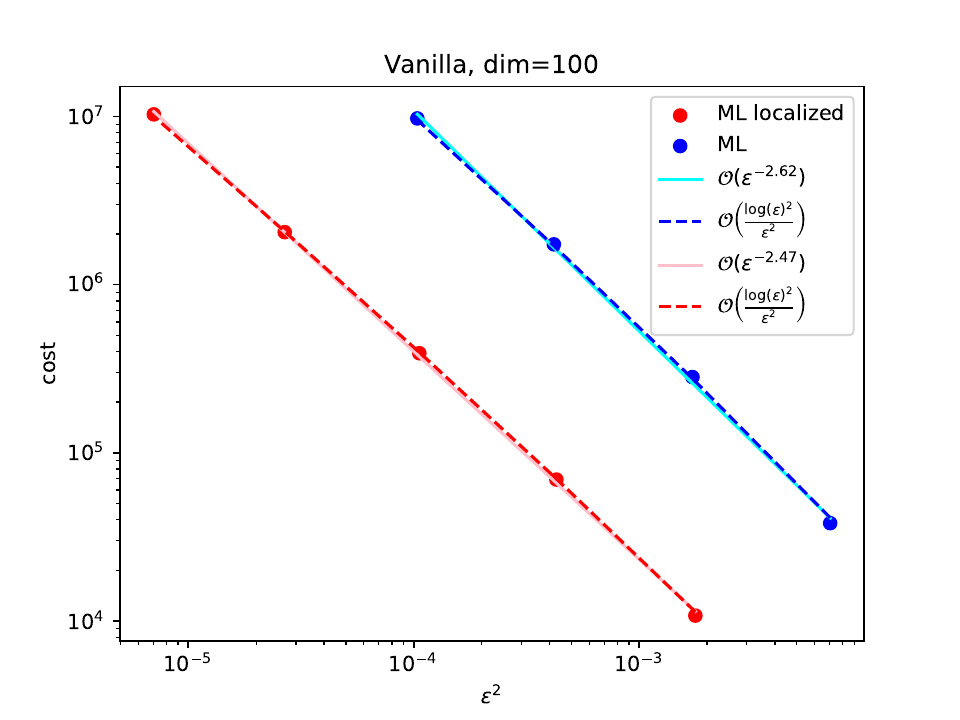}
\includegraphics[width=0.49\textwidth]{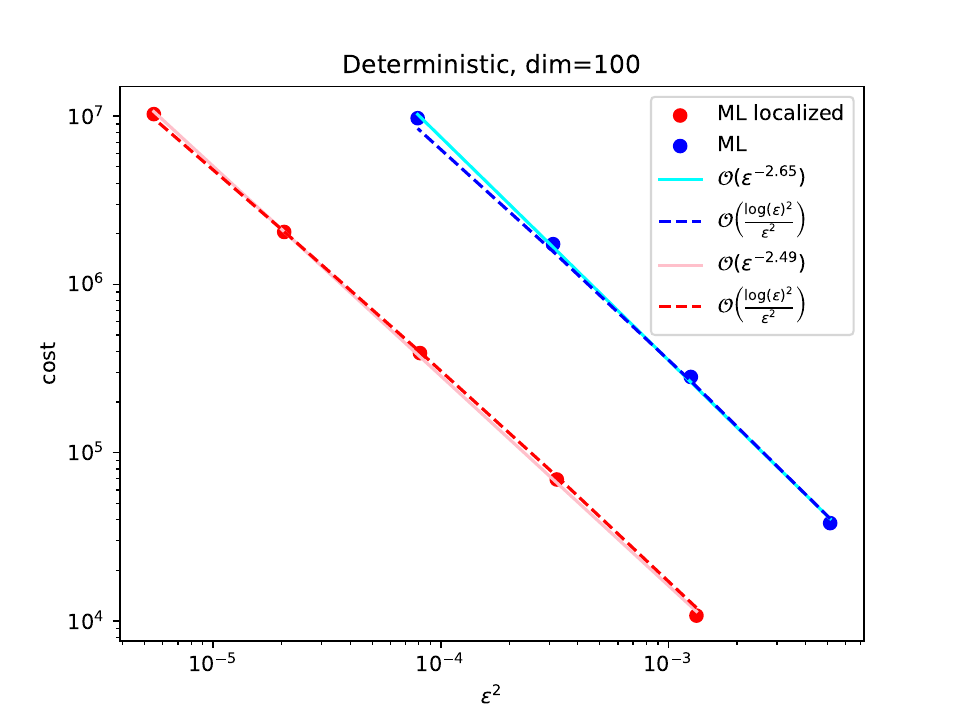}
 \caption{Computational complexity of plot for the Vanilla and Deterministic methods for dimension $d_x=$100. The error $\epsilon^2$ is the MSE.}
 \label{fig:local_funct1}
\end{figure}

\begin{figure}[h!]
\centering
\includegraphics[width=0.49\textwidth]{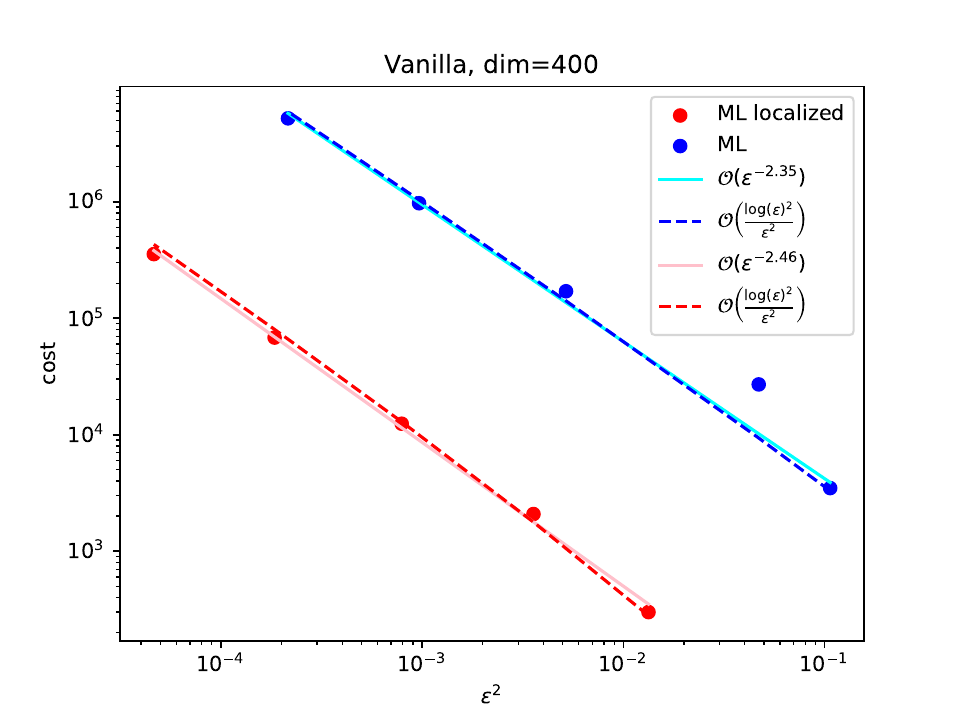}
\includegraphics[width=0.49\textwidth]{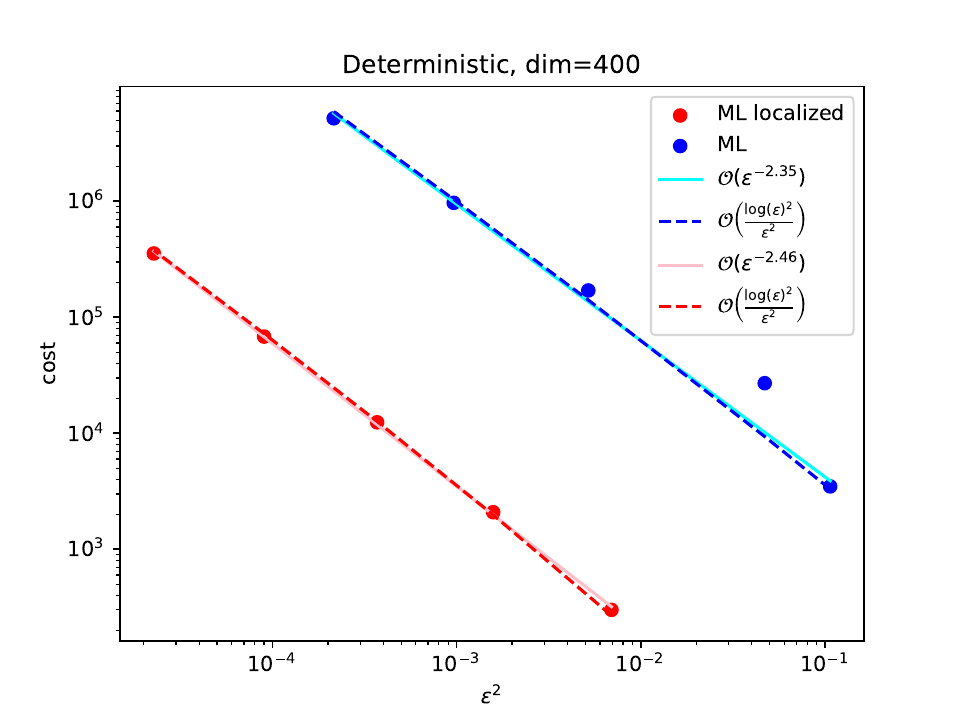}
 \caption{Computational complexity of plot for the Vanilla \textbf{(L-F1)} and Deterministic \textbf{(L-F2)} methods for dimension $d_x=$400. The error $\epsilon^2$ is the MSE.}
 \label{fig:local_funct2}
\end{figure}

This model is based on Equations \eqref{eq:data} - \eqref{eq:signal}, where each component of $X$ represents a point in an uniform 2D grid, thus $d_x=k^2$, for $k\in \mathbb{N}$. The mapping from the grid positions $(i,j)\in \{m\}_{m=0}^{k-1} \times \{m\}_{m=0}^{k-1} $ to the vector components is $d=\Omega(i,j)=ik+j$, the inverse mappings are denoted as $j=\Omega_2^{-1}(d)=d \text{ mod } k$, and $i=\Omega_1^{-1}(d)=[d-(d \text{ mod } k)]/k$. The drift matrix $A$ is chosen to allow direct interaction of the components only if they are within certain distance in the grid, such distance is measured by the euclidean metric, i.e., let $(p,q)\in  \{m\}_{m=0}^{k^2-1} \times \{m\}_{m=0}^{k^2-1}$ index the components of the drift $A$, then
\begin{align*}
    A^{(p,q)}=0, \quad \text{if} \quad  \sqrt{(i_p-i_q)^2+(j_p-j_q)^2}> r, 
\end{align*}
where $(i_p,j_p)$ and  $(i_q,j_q)$ are the respective positions in the grid of the components $p
$ and $q$. For simplicity, we choose the observation matrix $H\in\mathbb{R}^{d_y}$, $d_y=d_x$ proportional to the  diagonal matrix and  as well as the matrices $R_1$ and $R_2$. We use the parameters $r=1.5$ to allow interaction with at most 8 neighbors on the grid and $d=1.4$ as in \cite{WT20}. The final time of the computations is $T=10$ and we use two dimensions, $d_x=100$ and $d_x=400$. The MSE is computed using the KBF as the solution. From Figures \ref{fig:local_funct1} and \ref{fig:local_funct2} we can see the errors of both the localized and the standard multilevel EnKBF, for both the vanilla and the deterministic versions, and both $d_x=100$ and $d_x=400$. In the plots is evident that for a given error the cost of the localized version is smaller than the standard version (about $10^{-1}\sim 10^{-2}$ smaller). The rates are not exactly the same given the intricacies of the multilevel rates, nevertheless, they both fit the cost $\mathcal{O}(\frac{\log(\varepsilon^2)}{\varepsilon^2})$ at different points of the curve.

\begin{figure}[h!]
\centering
\includegraphics[width=0.49\textwidth]{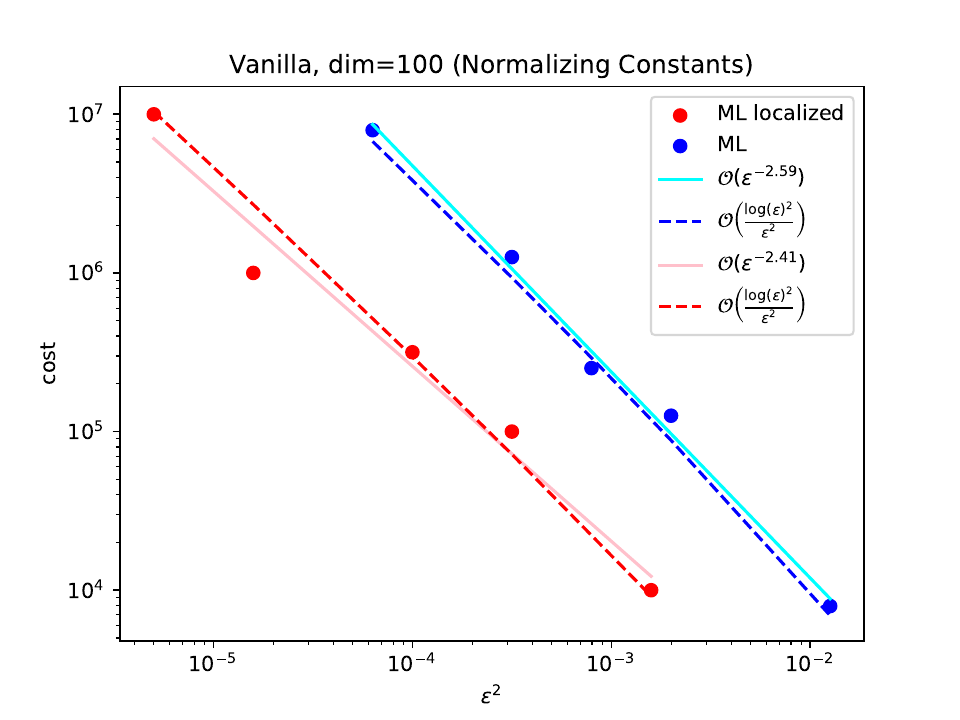}
\includegraphics[width=0.49\textwidth]{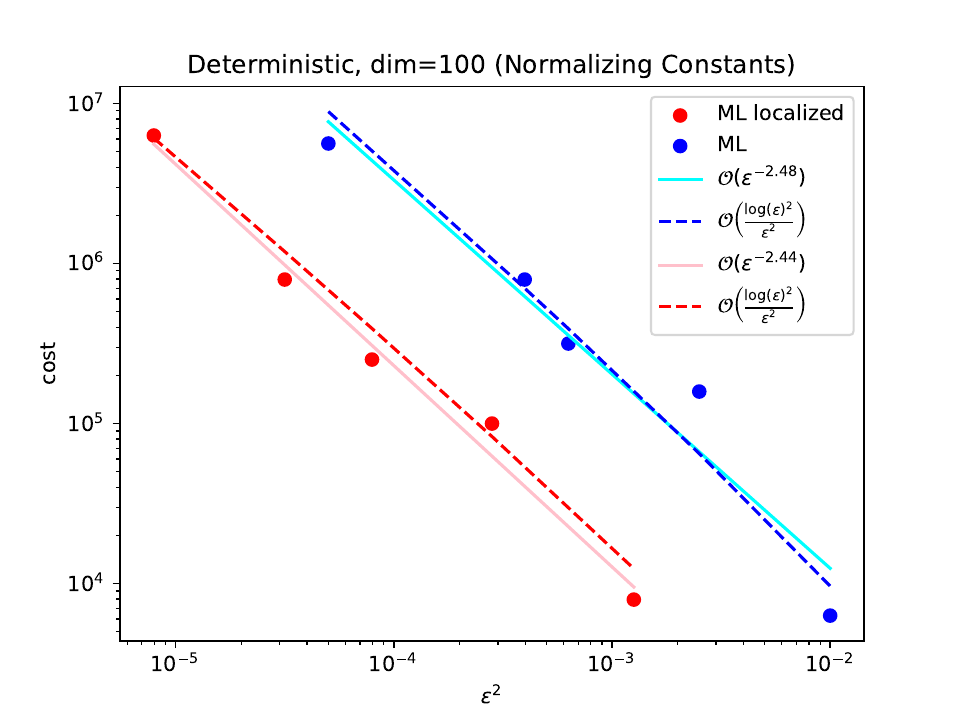}
 \caption{Computational complexity of normalizing constant plot for the Vanilla \textbf{(L-F1)} and Deterministic \textbf{(L-F2)} methods for dimension $d_x=$100. The error $\epsilon^2$ is the MSE.}
 \label{fig:local_funct1_nc}
\end{figure}

\begin{figure}[h!]
\centering
\includegraphics[width=0.49\textwidth]{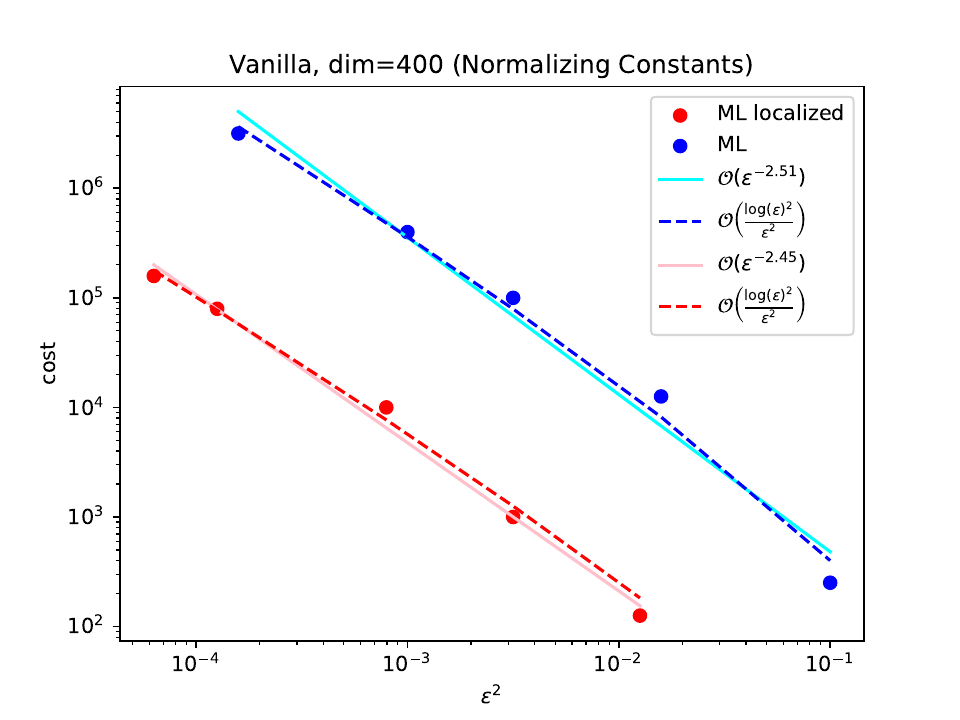}
\includegraphics[width=0.49\textwidth]{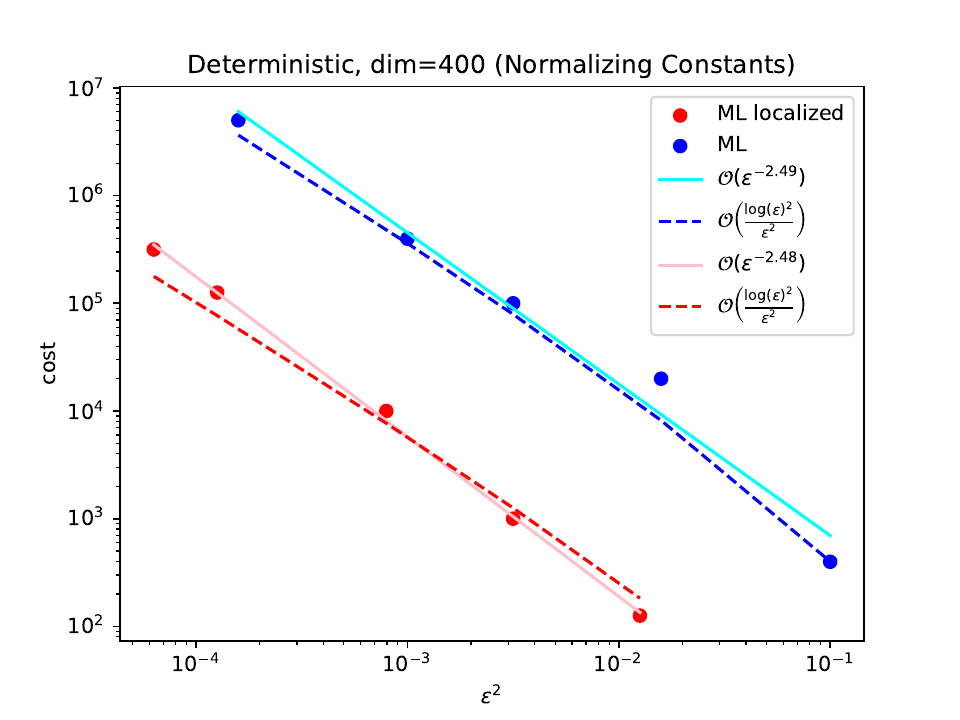}
 \caption{Computational complexity of normalizing constant plot for the Vanilla \textbf{(L-F1)} and Deterministic \textbf{(L-F2)} methods for dimension $d_x=$400. The error $\epsilon^2$ is the MSE.}
 \label{fig:local_funct2_nc}
\end{figure}

Finally we also consider a comparison for the estimation of the normalizing constants, which are presented in Figures 
\ref{fig:local_funct1_nc} and \ref{fig:local_funct2_nc}. As we can see, the rates obtained in those figures match exactly what we expect, which also coincides with Figures \ref{fig:local_funct1} and \ref{fig:local_funct2}.

\subsection{Parameter estimation}

Let us firstly assume that the model \eqref{eq:data} - \eqref{eq:signal} contains unknown parameters $\theta\in \Theta \subseteq \mathbb{R}^{d_\theta}$. To account for this, we rewrite the normalizing constant \eqref{eq:nc} with the additional subscript $\theta$ as $\overline{Z}_{t,\theta}(Y)$. To estimate these parameters we focus on maximum likelihood inference and stochastic gradient
methods that are performed in an online manner. In particular we follow a recursive maximum likelihood (RML), which has been discussed in \cite{BCJKR20, DDS10, PDS11} within the context of sequential Monte Carlo (or particle filtering) approximations. Let 
$$
U_{t,t+1,\theta}(Y) := \log \frac{\overline{Z}_{t+1,\theta}(Y) }{\overline{Z}_{t,\theta}(Y) }.
$$
RML relies on the following update scheme at any time $t\in \mathbb{N}$:
\begin{align*}
\theta_{t+1} &= \theta_t + a_t \left(  \nabla_\theta \log \overline{Z}_{t+1,\theta_t}(Y)  - \nabla_\theta \log \overline{Z}_{t,\theta_t}(Y)  \right) \\
&= \theta_t + a_t \nabla_\theta~ U_{t,t+1,\theta_t}(Y),
\end{align*}
where $\{a_t\}_{t\in\mathbb{N}}$ is a sequence of positive real numbers such that we assume the usual Robbins--Munro conditions, i.e. $\sum_{t\in\mathbb{N}} a_t =\infty$ and $\sum_{t\in\mathbb{N}} a_t^2 < \infty$. Given an initial $\theta_0\in\Theta$, this formula enables us to update $\theta$ online as we obtain a new observation path in each unit time interval. Computing
the gradients in the above formula can be expensive (see e.g. \cite{BCJKR20}), therefore we use a gradient-free method that is based on some type of finite differences with simultaneous perturbation stochastic approximation (SPSA) \cite{JCS03,JCS92}. In a standard finite difference approach, one perturbs $\theta_t$ in the positive and negative directions of a unit vector $\textbf{e}_k$ (a vector of zeros in all directions except $k$ it is 1). This means evaluating $U_{t,t+1,\theta_t}(Y)$ $2d_\theta$--times. 

Whereas in SPSA, we perturb $\theta_t$ with a magnitude of $b_t$ in the positive and negative directions of a $d_\theta$-dimensional random vector $\Psi_t$. The numbers $\{b_t\}_{t\in \mathbb{N}}$ are a sequence of positive real numbers such that $b_t \to 0$, $\sum_{t\in \mathbb{N}} a_t^2/b_t^2<\infty$, and for $k\in\{1\cdots,d_\theta\}$, $\Psi_t(k)$ is sampled from a Bernoulli distribution with success probability $1/2$ and support $\{-1,1\}$. Therefore, this method requires only 2 evaluations of $U_{t,t+1,\theta_t}(Y)$ to estimate the gradient. The MLEnKBF-NC estimator, presented in Algorithm \ref{alg:MLEnKBF_NC}, will be used to estimate the ratio $U_{t,t+1,\theta_t}(Y)$. In Algorithm \ref{alg:param_est} we illustrate how to implement these approximations in order to estimate the model's static parameters.

\begin{algorithm}[h!]
\caption{Parameter Estimation: using \textcolor{black}{MLEnKBF-NC} and RML\textcolor{black}{-}SPSA}
\label{alg:param_est}
\begin{enumerate}
\item \textbf{Input:} Target level $L\in\mathbb{N}$, start level $l_*\in \mathbb{N}$ such that $l_*<L$, the number of particles on each level $\{N_l\}_{l=l_*}^L$, the number of iterations $M\in\mathbb{N}$, initial $\theta_0\in\Theta$, step size sequences of positive real numbers $\{a_t\}_{t\in\mathbb{N}}$, $\{b_t\}_{t\in\mathbb{N}}$ such that $a_t,b_t \to 0$, $\sum_{t\in\mathbb{N}} a_t=\infty$, $\sum_{t\in\mathbb{N}} a_t^2/b_t^2<\infty$, and initial ensembles $\{\tilde{\xi}_0^{i}\}_{i=1}^{N_{tot}} \stackrel{i.i.d.}{\sim} \mathcal{N}(\mathcal{M}_0,\mathcal{P}_0)$, where $N_{tot}=\sum_{l=l_*}^L N_l$.

\item \textbf{Iterate:} For $t \in \{0,\cdots,M-1\}$: 
\begin{itemize}
\item Set $\{\xi_0^{i,l_*}\}_{i=1}^{N_{l_*}}=\{\tilde{\xi}_0^{i}\}_{i=1}^{N_{l_*}}$, $\cdots$, $\{\xi_0^{i,L}\}_{i=1}^{N_L}=\{\tilde{\xi}_0^{i}\}_{i=N_{L-1}+1}^{N_L}$.
\item For $k\in \{1,\cdots,d_\theta\}$, sample $\Psi_t(k)$ from a Bernoulli distribution with \\ success probability $1/2$ and support $\{-1,1\}$. 
\item Set $\theta_t^+= \theta_t + b_{t+1} \Psi_t$ and $\theta_t^-= \theta_t - b_{t+1} \Psi_t$.
\item Run Algorithm \ref{alg:MLEnKBF_NC} twice, with $T=1$ and initial ensembles $\Big\{ \{\xi_0^{i,l_*}\}_{i=1}^{N_{l_*}},\cdots,\{\xi_0^{i,L}\}_{i=1}^{N_L} \Big\}$, \\ to generate the estimates $U_{t,t+1,\theta_t^+}(Y)$ and $U_{t,t+1,\theta_t^-}(Y)$.
\item Set for $k\in\{1,\cdots,d_\theta\}$,
\begin{align*}
\theta_{t+1}(k) &= \theta_t(k) + \frac{a_{t+1}}{2b_{t+1}\Psi_t(k)} \Big[ U_{t,t+1,\theta_t^+}(Y)-U_{t,t+1,\theta_t^-}(Y) \Big].
\end{align*}
\item Run the EnKBF up to time 1 under the new parameter $\theta_{t+1}$ with discretization level $L$ and initial ensembles $\{\tilde{\xi}_0^{i}\}_{i=1}^{N_{tot}}$ to return $\{\tilde{\xi}_1^{i}\}_{i=1}^{N_{tot}}$.
\item Set $\{\tilde{\xi}_0^{i}\}_{i=1}^{N_{tot}}=\{\tilde{\xi}_1^{i}\}_{i=1}^{N_{tot}}$.
\end{itemize}
\end{enumerate}
\end{algorithm}

We will now introduce our different models we work with. This includes a linear and two nonlinear models, namely, the stochastic Lorenz 96 model.

\begin{figure}[h!]
\centering
\hspace*{-1.8cm}\includegraphics[width=1.2\textwidth]{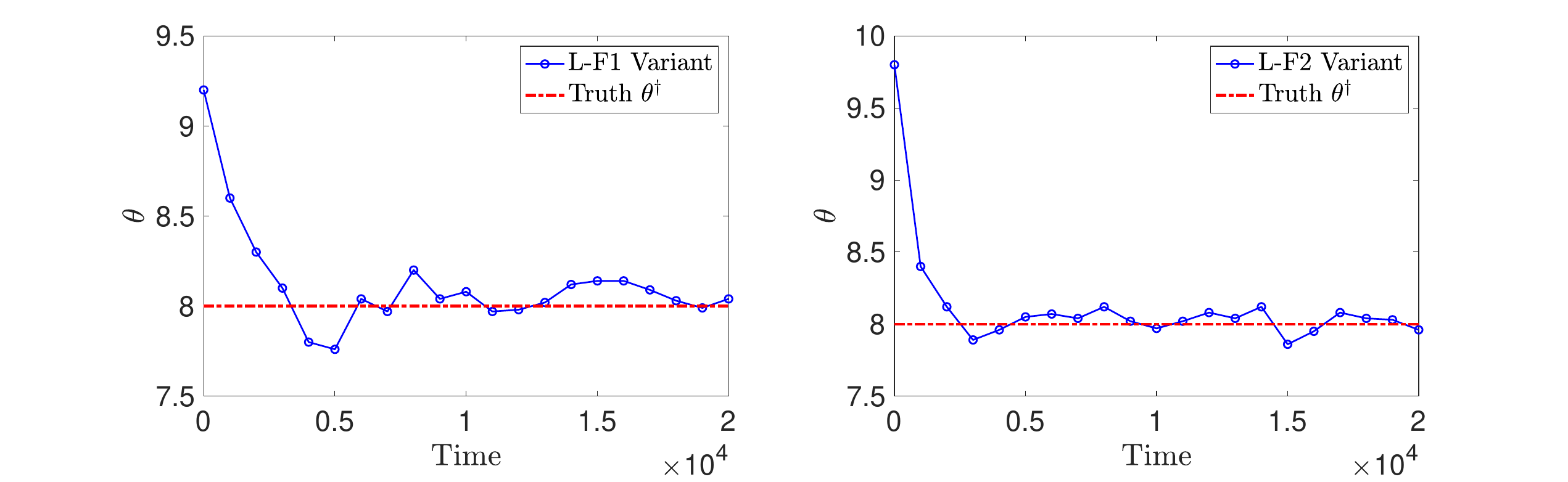}
 \caption{Computational complexity of plot for the Vanilla and Deterministic methods for dimension $d_x=$40. The error $\epsilon^2$ is the MSE.}
 \label{fig:param_est}
\end{figure}

Our final numerical experiment, will not be focused on testing the effective of the variance, or the MLMC rates, but instead 

test model will be the Lorenz 96 model \cite{ENL96} with $d_x=d_y=40$, which is a dynamical system designed to describe equatorial waves in atmospheric science. The stochastic Lorenz 96 model takes the form
\begin{align*}
dX_t &=   f(X_t) dt +  \textcolor{black}{Q}^{1/2} dW_t, \\
dY_t &=  C X_t dt +  \textcolor{black}{R}^{1/2}dV_t,  
\end{align*}
such that 
\begin{equation*}
f_i(X_t) = (X_t(i+1)-X_t(i-2))X_t(i-1)-X_t(i)+\theta,
\end{equation*}
where again $X_t(i)$ is the $i^{th}$ component of $X_t$, and we assume that $X_t(-1) = X_t(d_x-1)$, $X_t(0) = X_t(d_x)$ and that $X_t(d_x+1) = X_t(1)$. We specify our parameter values as $\textcolor{black}{Q}^{1/2} = \sqrt{2}~Id$ and $\textcolor{black}{R}^{1/2} = 0.5~ Id$. The parameter $\theta$ is the external force in the system, while $(X_t(i+1)-X_t(i-2))X_t(i-1)$ is the advection term and $-X_t(i)$ is the damping term. We set the target level to be $L=9$, and the start level $l_*=7$. In \textbf{(F1)} and \textbf{(F2)}, we specify the initial state as follows: We set $X_0(1) = 8.01$ and $X_0(i)=8$ for $1<i\leq d_x$. We specify the true underlying parameter as $\theta^* = 8$.
\\
Our final experiment is provided in Figure \ref{fig:param_est}, where we use Algorithm \ref{alg:MLEnKBF_NC}, where we provide the learning of the parameter $\theta$ for both localized multilevel EnKBFs. As we observe the learning time takes longer for the vanilla version due to the addition of the observational noise. However both methods learn the parameter after 5000 seconds, which remains consistent afterwards. This verifies the localized method can work for parameter estimation. Table \ref{table:1} presents the variance  of the 50 different runs compared to the non-localized versions. The reduction in variance is a further indication of the potential of localization techniques.

\begin{table}[h!]
\begin{center}
\begin{tabular}{ | m{5em} | m{2cm}| m{2cm} | } 
  \hline
\textbf{EnKBF Method} & \textbf{Running Mean} & \textbf{Running Variance} \\ 
  \hline
(F1) & 8.08 & 0.30 \\ 
  \hline
  (L-F1) & 8.05 & 0.19 \\ 
  \hline
(F2) & 8.07 & 0.25 \\ 
  \hline  
(L-F2) & 8.02 & 0.16 \\ \hline
\end{tabular}
\caption{Table comparing running mean and variances of different EnKBF algorithms.}
\label{table:1}
\end{center}
\end{table}

\section{Conclusion \& Discussion}
\label{sec:conc}

Localization is a powerful technique which adds considerable benefits to the ensemble Kalman filter within data assimilation. It does so by reducing spurious correlations, which in turn improves on stability and accuracy for state and parameter estimation. For localization one can rely on the assumption that the ensemble members are less than the dimension, i.e. $N< d_x$ (where we assume here that $d_x=d_y$). In this work we considered applying this technique to the continuous-time filtering algorithm of the ensemble Kalman-Bucy filter, coupled with multilevel Monte Carlo. The later method is a well-known methodology to reduce the cost to attain a particular order of mean square error. Therefore our focus was the development of a computational sound method. Our algorithm proposed was the LMLEnKBF for which we tested this on a range of problems, both for parameter and state estimation. The former includes a linear-dynamical and nonlinear model such as the Lorenz 96 model, and finally a linear-Gaussian model for state estimation. Our results showcase improvements by remaining competitive with the  non-localized MLEnKBF, whilst improving on computational time and stability. The former was the case where we able to see the MSE-to-cost rate being $\mathcal{O}(\epsilon^2 \log(\epsilon)^2)$. While for the latter, this was seen through both MSE and variance plots of our Monte Carlo estimators.

For future work there are various paths we can improve on, based on this work. Firstly it would be of interest to apply these techniques ti parameter estimation problems arising in partial differential equations, which includes turbulence flow models. This could involve the use of ensemble Kalman methods \cite{CCS21,CIRS18,CT22}, which has already been tested, in a discrete-time fashion with covariance localization \cite{TM23}. Another direction of interest is to see if the techniques of domain localization \cite{JNA11} could be applied. This is again particularly useful in the case of PDEs, therefore careful inspection is required to enable this extension, as here we consider stochastic and ordinary differential equations. Finally a more obvious research directions, or directions, is that of applying to this to nonlinear models where one can attain MSE-to-cost rates. This goes beyond the current work, as it is based on the work of Chada et al. \cite{CJY20}. This would also promote new potential theory beyond the work of this paper, which could provide a more complete understanding of multilevel and localized EnKBFs.


\section*{Acknowledgements}
NKC is supported by an EPSRC-UKRI AI for Net Zero Grant: ``{Enabling CO2 Capture And Storage Projects Using AI}". NKC is also supported by a City University of Hong Kong startup grant: 7200809. NKC also thanks Miguel Alvar\'{e}z for helpful discussions on the implementation.


\begin{thebibliography}{99}


\bibitem{ACJ22}
{\sc \'{A}lvarez}, M., {\sc Chada}, N. K. C. \& {\sc Jasra}, A.~(2022)
\newblock{Unbiased Estimation of the vanilla and deterministic ensemble Kalman--Bucy filters.}
\newblock{arXiv preprint arXiv:2208.03947, 2022}.

\bibitem{JLA07}
{\sc Anderson}, J. L.~(2007) 
Exploring the need for localization in ensemble data assimilation using a hierarchical ensemble filter. 
\newblock{\em Physica D: Nonlinear Phenomena}, 230(1), 99--111.

\bibitem{AA99}
{\sc Anderson}, J. L. \& {\sc Anderson}, S. L.~(1999)
A Monte Carlo implementation of the nonlinear filtering problem to produce ensemble assimilations and forecasts. 
\newblock{\em Mon. Wea. Rev.}, 126, 2741--2758.


\bibitem{BC09}
{\sc Bain}, A. \& {\sc Crisan}, D.~(2009)
\newblock{\em Fundamentals of Stochastic Filtering}.
\newblock{Springer, New York}.


\bibitem{BR10}
{\sc Bergemann}, K. and {\sc Reich}, R.~(2010) 
A localization technique for ensemble Kalman filters.
\newblock{\em Quart. J. Roy. Meteor. Soc.}, 136(648):701--707.


\bibitem{BCJKR20}
A. Beskos, D. Crisan, A. Jasra, N. Kantas and H. Ruzayqat. 
Score-based parameter estimation for a class of continuous-time state space models. 
\newblock{\em SIAM J. Sci. Comp. (to appear)}, 2021.

\bibitem{BD20}
A. N. Bishop and P. Del Moral.
\newblock{On the mathematical theory of ensemble (linear-gaussian) Kalman--Bucy filtering.}
arXiv preprint arXiv:2006.08843, 2020.



\bibitem{RSB65}
{\sc Bucy}, R. S.~(1965)
Nonlinear filtering theory. \emph{IEEE Trans. Automat. Control}, 10, {198}.


\bibitem{CJY20}
{\sc Chada}, N. K.,  {\sc Jasra},  A. \& {\sc Yu}, F.~(2022).
Multilevel ensemble Kalman--Bucy filters. \emph{SIAM/ASA Journal on Uncertainty Quantification}, 9(2), 763--787.

\bibitem{CCS21}
{\sc Chada}, N. K., {\sc Chen}, Y. \& {\sc Sanz-Alonso}, D.~(2021)
Iterative ensemble Kalman methods: A unified perspective with some new variants.
\newblock{\em Foundations of Data Science}, 3(3), 331--369.


\bibitem{CIRS18}
{\sc Chada}, N. K., {\sc Iglesias}, M. A., {\sc Roininen}, L., \& {\sc Stuart}, A. M.~(2018)
\newblock{Parameterizations for ensemble Kalman inversion.}
\newblock{\em Inverse Problems}, 34 (5), 055009.


\bibitem{CST20}
{\sc Chada}, N. K., {\sc Stuart}, A. M. \& {\sc Tong}, X. T.~(2022)
\newblock{Tikhonov regularization within ensemble Kalman inversion.}
\newblock{\em SIAM Journal on Numerical Analysis}, 58 (2), 1263--1294.

\bibitem{CT22}
{\sc Chada}, N. K., \& {\sc Tong}, X. T.~(2022)
\newblock{Convergence acceleration for ensemble Kalman inversion in nonlinear settings.}
\newblock{\em Math. of Comp.}, 91 (335), 247--1280.



\bibitem{CDJ21}
{\sc Crisan}, D., {\sc Del Moral}, P., {\sc Jasra}, A. \& {\sc Ruzayqat}, H.~(2022) 
Log-normalization constant estimation using the ensemble Kalman-Bucy filter with application to high-dimensional models.
\newblock{\em Adv. Appl. Probab. (to appear)}.


\bibitem{CR11}
{\sc Crisan}, D. \& {\sc Rozovskii}, B.~(2011)
\newblock{\em The Oxford Handbook of Nonlinear Filtering}.
Oxford University Press, Oxford, 2011.


\bibitem{DDS10}
P. Del Moral, A. Doucet and S. S. Singh.
Forward smoothing using sequential Monte Carlo.
arXiv preprint arxiv:1012.5390, 2010.

\bibitem{DT18}
{\sc Del Moral}, P.  \& {\sc Tugaut},  J.~(2018).
On the stability and the uniform propagation of chaos properties of ensemble Kalman--Bucy filters.
{\em Ann. Appl. Probab.}, {28}, 790--850.


\bibitem{GE09}
{\sc Evensen}, G.~(2009)
\newblock {\em Data Assimilation: The Ensemble Kalman Filter}. {Springer}.


\bibitem{GE94}
{\sc Evensen}, G.~(1994)
\newblock{Sequential data assimilation with a nonlinear quasi-geostrophic model using Monte Carlo methods to forecast error statistics}.
\newblock{\em Journal of Geophysical Research: Oceans},  {99}, 10143--10162.


\bibitem{FT17}
{\sc Fossum}, K. \& {\sc Mannseth}, T.~(2017)
Coarse-scale data assimilation as a generic alternative to localization.
{\em Computational Geosciences}, {21}(1), 167-186.


\bibitem{GC99}
{\sc Gaspari}, G. \& {\sc Cohn}, S. E.~(1999) 
Construction of correlation functions in two and three dimensions.
{\em Q. J. R. Meteorol. Soc}. {125} 723--57.




\bibitem{GM98}
{\sc Gelman}, A. \& {\sc Meng}, X. L.~(1998) 
 Simulating normalizing constants: From importance sampling to bridge sampling to path sampling. 
 \newblock{\em Statistical science}, 163--185.



\bibitem{MBG08}
{\sc Giles}, M. B.~(2008)
Multilevel Monte Carlo path simulation. 
\newblock{\em Op. Res.}, {56}, 607--617.


\bibitem{MBG15}
{\sc Giles}, M. B.~(2015)
\newblock{Multilevel Monte Carlo methods.}
\newblock{\em  Acta Numerica}, {24}, 259--328.



\bibitem{HWS01}
{\sc Hamill}, T. M., {\sc Whitaker}, J. S. \& {\sc Snyder}, C.~(2001)
Distance-dependent filtering of background error covariance estimates in an ensemble Kalman filter. 
{\em Mon. Wea. Rev}., {129}(11), 2776--2790.



\bibitem{HLT16}
{\sc Ho{e}l}, H., {\sc Law}, K. J. H. \& R. {\sc Tempone}, R.~(2016)
\newblock{Multilevel ensemble Kalman filtering}.
\newblock{\em SIAM J. Numer. Anal.}, {54}(3), 1813--1839.


\bibitem{HM98}
{\sc Houtekamer}, P. L. \& {\sc Mitchell}, H. L.~(1998)
Data assimilation using an ensemble Kalman filter technique. {\em Mon. Wea. Rev}., {126}, 796--811.


\bibitem{HM01}
{\sc Houtekamer}, P. L. \& {\sc Mitchell}, H. L.~(2001).
 A sequential ensemble Kalman filter for atmospheric data assimilation.  {\em Mon. Wea. Rev}., {129}(1), 123--137.

\bibitem{AJ70}
{\sc Jazwinski}, A.~(1970).
{\em Stochastic processes and filtering theory}.  Academic Pr.


\bibitem{JNA11}
{\sc Janjic}, T., {\sc Nerger}, L., {\sc Albertlla}, A., {\sc Schroter}, J. \& {\sc Skachoko}, S.~(2011) 
On domain localization in ensemble-based Kalman filter algorithms. 
\newblock{\em Mon. Wea. Rev.}, 139(7):2046--2060.


\bibitem{ACL86}
{\sc Lorenc}, A. C.~(1986). 
Analysis methods for numerical weather prediction. 
\newblock{\em Quart. J. Roy. Meteor. Soc.}, 122, 1177--1194.

\bibitem{ENL96}
{\sc Lorenz}, E.N.~(1996) Predictability: A problem partly solved. 
\newblock{\em Proc. ECMWF Seminar on predictability}, 1, 1--18.


\bibitem{MW06} 
{\sc Majda}, A. \& {\sc Wang}, X.~(2006) 
\newblock{\em Non-linear Dynamics and Statistical Theories for Basic Geophysical Flows,} Cambridge University Press.


\bibitem{NBF22}
{\sc Nezhadali}, M., {\sc Bhakta}, T., {\sc Fossum}, K. \& {\sc Mannseth}, T.~(2022)
Iterative multilevel assimilation of inverted seismic data.
{\em Computational Geosciences}, {26}(2), 241--262.


\bibitem{ORL08} 
{\sc Oliver}, D., {\sc Reynolds}, A. C. \& {\sc Liu}, N.~(2008)
\newblock{ \em Inverse Theory for Petroleum Reservoir Characterization and History Matching}. Cambridge University Press, 1st edn.


\bibitem{OHS04}
{\sc Ott}, E., {\sc Hunt}, B. R., {\sc Szunyogh}, I., {\sc Zimin}, A. V., {\sc Kostelich}, E. J., {\sc Corazza}, M., {\sc Kalnay}, E., {\sc Patil}, D. J. \& {\sc Yorke}, J. A.~(2004)
A local ensemble Kalman filter for atmospheric data assimilation. {\em Tellus A}, {56}, 415--428.

\bibitem{PDS11}
G. Poyiadjis, A. Doucet and S. S. Singh.
 Particle approximations of the score and observed information matrix in state space models with application to parameter estimation. 
 {\em Biometrika} 98: 65--80, 2011.

\bibitem{RJP18}
{\sc Rischard}, M., {\sc Jacob}, P. E. \& {\sc Pillai}, N.~(2018)
Unbiased estimation of log normalizing constants with applications to Bayesian cross-validation.
arXiv preprint arxiv:1810.01382, 2018.


\bibitem{RJS15}
{\sc Roh}, S.,  {\sc Jun}, M., {\sc Szunyogh}, I. \& {\sc Genton}, M. G.~(2015)
Multivariate localization methods for ensemble Kalman filtering.
{\em Nonlin. Processes Geophys. Discuss.}, {2}, 833--863.


\bibitem{RCJ21}
{\sc Ruzayqat}, H., {\sc Chada}, N. K. C. \& {\sc Jasra}, A.~(2022)
\newblock{Multilevel estimation of normalization constants using the ensemble Kalman--Bucy filter.}
\newblock{\em Statistics and Computing} {32}(38).


  \bibitem{WRS18}
{\sc de Wiljes}, J. , {\sc Reich} S. \& {\sc Stannat} W.~(2018).
 \newblock{Long-time stability and accuracy of the ensemble Kalman--Bucy filter for fully observed processes and small measurement noise.} 
 \newblock{\em SIAM J. Appl. Dyn. Syst.}, {17}(2), 1152--1181.


\bibitem{WT20}
{\sc de Wiljes}, J. \& {\sc Tong}, X. T.~(2020).
Analysis of a localised nonlinear Ensemble Kalman Bucy Filter with complete and accurate observations.
\newblock{\em Nonlinearity}., {33}(9), 4752--4782.


\bibitem{XTT18}
{\sc Tong}, X. T.~(2018).
Performance analysis of the localized ensemble Kalman filter. {\em J. Nonlinear Sci.}, {28} 1397--442.


\bibitem{TM23}
{\sc Tong}, X. T. \& {\sc Morzfield}, M.~(2023).
Localizend ensemble Kalman Inversion. {\em Inverse Problems}, {39}(6).




\end{thebibliography}
\end{document}